\documentclass[11pt]{article} 
\usepackage{amsfonts,latexsym}
\usepackage{epsf}
\usepackage[scanall]{psfrag}
\usepackage{graphicx}

\newcommand{\Sp}{\mathbb S}  
\newcommand{\R}{\mathbb R}
\newcommand{\RR}{\mathbb R}  

\newcommand{\N}{\mathbb N}  
\newcommand{\C}{\mathbb C}

\newcommand{\calC}{\mathcal C}
\newcommand{\calN}{\mathcal N}
\newcommand{\calL}{\mathcal L}
\newcommand{\calG}{\mathcal G}

\newcommand{\del}{\partial}  

\newcommand{\D}{{\mathcal D}}
\newcommand{\calD}{{\mathcal D}}
\newcommand{\e}{\varepsilon}

\newcommand{\Si}{\Sigma}

\newcommand{\ST}{\Sigma_{T}}
\newcommand{\SR}{\Sigma^{R}}

\newcommand{\g}{\gamma}
\newcommand{\gc}{\gamma_{c}}
\newcommand{\gt}{\gamma_{T}}
\newcommand{\Gt}{{\Gamma}_{T}}

\newcommand{\mut}{\mu_{T}}
\newcommand{\tmut}{{\tilde\mu}_{T}}

\newcommand{\pt}{\psi_{T}}
\newcommand{\tpt}{{\tilde \psi}_{T}}

\newcommand{\olg}{\overline{\gamma}}

\newcommand{\tgT}{\tilde{\gamma}_T}
\newcommand{\af}{\mbox{\tiny AE}}
\newcommand{\ah}{\mbox{\tiny AH}}

\newcommand{\di}{\hbox{div}}
\newcommand{\dive}{{\mbox{div\,}}}
\newcommand{\tr}{{\mbox{tr\,}}}

\newtheorem{theorem}{Theorem}  
\newtheorem{lemma}{Lemma}
\newtheorem{proposition}{Proposition}  
\newtheorem{corollary}{Corollary}
\newtheorem{definition}{Definition}  
\newtheorem{remark}{Remark}
 
\begin{document} 
 
\title{Gluing and wormholes for the Einstein constraint equations}

\author{James Isenberg\thanks{Supported by the NSF under Grant PHY-0099373}  
\\ University of Oregon  \and
Rafe Mazzeo\thanks{Supported by the NSF under Grant DMS-9971975 and 
at MSRI by NSF grant DMS-9701755}  
\\ Stanford University \and 
Daniel Pollack\thanks{Supported by the NSF under Grant DMS-9704515}  
\\ University of Washington}

\date{July 23, 2002}

\maketitle

\begin{abstract}
We establish a general gluing theorem for constant mean curvature
solutions of the vacuum Einstein constraint equations. This allows
one to take connected sums of solutions or to glue a handle
(wormhole) onto any given solution. Away from this handle region,
the initial data sets we produce can be made as close as 
desired to the original initial data sets. These constructions
can be made either when the initial manifold is compact or
asymptotically Euclidean or asymptotically hyperbolic, with
suitable corresponding conditions on the extrinsic curvature.
In the compact setting a mild nondegeneracy condition is
required. In the final section of the paper, we list a
number ways this construction may be used to produce new
types of vacuum spacetimes.
\end{abstract}

\section{Introduction} 
\label{se:1}

\subsection{The constraint equations and surgery}
The Einstein equations for the gravitational field on a Lorentzian 
four-manifold $Z$ form, modulo diffeomorphisms, 
a locally well-posed hyperbolic system \cite{CB},
and vacuum solutions may be obtained from a set of Cauchy data on a 
three-dimensional spacelike slice $\Sigma \subset Z$. A vacuum initial 
data set for this problem consists of two symmetric 
tensors $\gamma$ and $\Pi$ on 
$\Sigma$, which correspond respectively to the induced metric and second 
fundamental form on $\Sigma$. Thus $\gamma$ is positive definite while $\Pi$ 
is at least apparently unrestricted. However, in order for the Einstein
evolution to exist, at least for a short time, it is necessary and 
sufficient \cite{CB} that these tensors satisfy the constraint equations:
\begin{eqnarray}
\dive \Pi - \nabla \tr \Pi & = & 0  \label{eqn1}\\
R_\gamma - |\Pi|^2_\gamma + (\tr \Pi)^2 & = & 0 \label{eqn2}
\end{eqnarray}
Here $\nabla$ is the Levi-Civita connection for $\gamma$, $R_\gamma$
its scalar curvature, and the divergence, $\dive_\gamma$, trace, $\tr_\gamma$
and norm squared of $\Pi$, $|\Pi|^2_\gamma = \Pi_{ab}\Pi_{cd}\gamma^{ac}
\gamma^{bd}$, are all computed with respect to $\gamma$. 

Having fixed the three-manifold $\Sigma$, it is not apparent that solutions 
$(\gamma,\Pi)$ of these constraint equations exist, or how to find them. 
However, there is a systematic procedure which generates solutions in a 
large number of cases, which we review now. Let us begin with a pair of 
tensors $(\gamma,\Pi)$, which do not necessarily satisfy the constraint 
equations, but which have the property that the mean curvature 
$\tau = \tr_\gamma \Pi$ is constant. The method we are describing relies
crucially on this assumption. Decompose $\Pi$ into its trace-free 
and pure trace parts:
\begin{equation}
\Pi = \mu + \frac{\tau}{3}\gamma.
\label{eq:splpi}
\end{equation}
The first of the constraint equations (\ref{eqn1}) requires that $\dive \mu 
= 0$, since $\tau$ is constant, so that $\mu$ is what is usually called a 
transverse-traceless (TT) tensor field. The idea is to modify the pair 
$(\gamma,\Pi)$ by changing $\gamma$ and the {\it trace-free part} $\mu$ 
of $\Pi$ by a conformal factor; by a judicious choice of this factor,
the new tensors will satisfy the constraint equations. More specifically, 
for any $\phi > 0$ we set
\begin{eqnarray}
\tilde{\gamma}  & =  & \phi^4\gamma, \label{eq:conf1} \\
\tilde{\Pi} \ \ = \ \ \tilde{\mu} + \frac{\tau}{3} \tilde{\gamma}
& = & \phi^{-2}\mu + \frac{\tau}{3}\phi^{4}\gamma. \label{eq:conf2}
\end{eqnarray}
The factor $\phi^{-2}$ in $\tilde{\mu}$ is the only one for which 
$\tilde{\mu}$ is divergence-free with respect to $\tilde{\gamma}$.
(Note that $\tilde{\Pi}$ and $\tilde{\mu}$ as they appear in
(\ref{eq:conf2}) are covariant tensors; the factor $\phi^{-10}$ replaces
$\phi^{-2}$ if we work with 
$\tilde{\Pi}$ and $\tilde{\mu}$ in contravariant form.)
This new pair $(\tilde{\gamma},\tilde{\Pi})$ satisfies the constraint 
equations provided the conformal factor $\phi$ satisfies the Lichnerowicz 
equation 
\begin{equation}\label{eqn3}
\Delta\phi - \frac{1}{8}R\phi + \frac{1}{8}|\mu|^2\phi^{-7}- 
\frac{1}{12}\tau^2\phi^5 = 0.
\end{equation}
Here the Laplacian, scalar curvature and norm-squared of $\mu$
are all computed relative to $\gamma$. By well-known analytic techniques 
one may find solutions of this semilinear equation, and so altogether 
this procedure produces many admissible pairs of tensors satisfying the 
constraint equations. In effect, the constancy of the mean curvature 
$\tau$ decouples the equations (\ref{eqn1}), (\ref{eqn2}), reducing
them to the facts that $\mu$ is transverse-traceless and $\phi$ satisfies
the Lichnerowicz equation. These formulations are equivalent and so
we label initial data sets in one of three equivalent ways: as 
$(\gamma,\Pi)$ if these tensors satisfy the constraint equations, or as 
$(\gamma,\Pi,\phi)$ if $\phi$ is the conformal factor which alters
$(\gamma,\Pi)$ to tensors satisfying the constraint equations, 
or finally (suppressing $\tau$) as $(\gamma,\mu,\phi)$. When $\phi$ 
is omitted in this notation, then $\phi = 1$ is implied.

When $\Sigma$ is compact, complete existence results are available
in this CMC case \cite{I1}; analogous results are known when $\Sigma$ 
is either asymptotically flat \cite{Ca1}, \cite{CBY}, \cite{CCB}, \cite{COM} 
or asymptotically hyperboloidal 
\cite{ACF}, \cite{AC}. In the non-CMC case, when $\tau$ is no longer constant, 
only partial results have been obtained \cite{IM}, \cite{CB2},
\cite{CIY}, \cite{IP} but this direction
warrants greater attention since the existence of constant mean curvature 
slices is apparently a restrictive assumption in relativity \cite{Bar2}, 
\cite{Br}.

In this paper we address the following question. Suppose we are given 
a three-manifold $\Sigma$ along with an initial data set $(\gamma,\Pi)$ 
on it which satisfy the constraint equations. Then is it possible to modify 
$\Sigma$ by `surgery' to obtain a new, topologically distinct manifold 
$\Sigma'$, and to find tensors $\gamma'$ and $\Pi'$ on $\Sigma'$ which 
themselves satisfy the constraint equations?  Since we can find solutions 
by applying the procedure above 
to $\Sigma'$ in the first place, the more precise question we propose to 
study is whether it is possible to perform the surgery on a geometrically 
small subset of $\Sigma$ and to find solutions $(\gamma',\Pi')$ which are
very close to the initial solutions $(\gamma,\Pi)$ away from this subset?

Let us phrase this more concretely. There are many different types of 
surgeries possible on higher dimensional manifolds, and really only
two types in three dimensions. One is akin to Dehn surgery, while the
other, on which we focus, consists of `adding a handle' in the following 
sense. Choose any two points $p_1, p_2 \in \Sigma$ and 
let $B_1$ and $B_2$ be small (metric) balls centered at these points. 
Excise these balls to obtain a manifold $\Sigma''=\Sigma \setminus 
(B_1\cup B_2)$. The boundary of $\Sigma''$ consists of two copies of $S^2$,
the same as the product of an interval with a two-sphere, $I \times S^2$.
Identifying these two manifolds along this common boundary yields 
a new manifold 
\[
\widehat{\Sigma} = \Sigma'' \bigsqcup_{\{\pm 1\} \times S^2} I \times S^2.
\]
Informally, we propose to restrict the original initial data set $(\gamma,
\Pi)$ to $\Sigma''$, modify this data slightly, and then find a one-parameter
family of extensions $(\gamma_\e,\Pi_\e)$ of this data to $\widehat{\Sigma}$, 
$\e\in (0,\e_0)$, each of which also satisfy the constraint equations. 
The parameter $\e$ measures the size of the handle. We shall do this
in such a way that on any compact set $K$ away from the handle,
$(\gamma_\e,\Pi_\e)$ converges to $(\gamma,\Pi)$ as some power of $\e$. 
The data on the handle will converge to zero. 
In the compact setting, the only condition needed 
to make this procedure work is a very mild nondegeneracy condition
as well as the assumption that $\Pi\not \equiv 0$; the precise statement 
is contained in the main theorem below. In particular, we 
cannot handle compact `time-symmetric slices', 
i.e.\ those with $\Pi \equiv 0$,
in which case the constraint equations reduce to the vanishing of scalar 
curvature, and known topological obstructions (see \cite{RS} for
example) prevent this sort of 
construction from working in general then. 

There are two main cases of this construction. In the first, $\Sigma$
consists of two components, $\Sigma_1$ and $\Sigma_2$, and the two points 
$p_1, p_2$ lie in these separate components. Then we are producing
solutions of the constraint equations on the connected sum
$\Sigma_1 \# \Sigma_2$ of these two components. In the other case,
$\Sigma$ is connected, and this construction may be regarded as 
demonstrating that a `wormhole' can be added to most solutions of
the constraint equations. Both cases are of clear physical interest.

The analytic techniques used in the proof here are in the spirit of many 
other gluing theorems for a variety of other geometric structures. Gluing 
theorems are fundamental in gauge theory, and similar techniques have been 
developed and applied to metrics of constant scalar curvature \cite{MPU}, 
\cite{MPa1}, minimal and constant mean curvature surfaces \cite{K1}, 
\cite{MPa}, \cite{MPP}, 
and a variety of other geometric problems. Another recent result 
using the same circle of ideas, and relevant to this area of general 
relativity, is contained in \cite{Cor}; this paper proves the existence
of time-symmetric asymptotically flat solutions to the constraint equations 
(i.e. scalar flat asymptotically Euclidean metrics) which evolve to 
spacetimes which are identically Schwarzchild near spatial infinity. 
The technique common to many of these papers consists of a two-step procedure.
First a one-parameter family of approximate solutions to any one of these 
problems is constructed. These are chosen 
so that the error measuring the discrepancy of these approximate solutions 
from exact solutions tends to zero, and so the strategy is to perturb
these approximate solutions to exact solutions using a contraction
mapping or implicit function theorem argument. The complication here
is that as the parameter $\e$ converges to zero, the geometry is
degenerating and some care must be taken to control the linearization
of the nonlinear equation associated to the geometric problem.
A slightly different and ultimately simpler approach is contained in 
\cite{MPa}, \cite{MPP}; however, we shall follow the earlier and 
more familiar route here because in the present context either method 
requires approximately the same amount of work. 

In the remainder of this section we set up some notation, give some
definitions, state precise versions of our main results, and
provide a more detailed guide to the rest of the paper.
 
\subsection{Statement of main results} 
\label{se:1.2}

We fix a three-manifold $\Sigma$ along with a conformal class $[\gamma]$
on $\Sigma$. For any two points $p_1,p_2$ on this manifold, we refer to 
the triple $([\gamma],p_1,p_2)$ as a marked conformal structure on 
$\Sigma$. We have already alluded to a nondegeneracy condition required 
for the construction:

\begin{definition}\label{nondegen} 
A marked conformal structure $([\gamma],p_1,p_2)$ on a compact manifold 
$\Sigma$ is {\bf nondegenerate} if either of the following situations
is true:
\begin{itemize}
\item $\Sigma$ is connected and there are no nontrivial conformal Killing 
vector fields on $\Sigma$ which vanish at either $p_1$ or $p_2$, or
\item The points $p_1$ and $p_2$ lie on different components of
$\Sigma$ and there are no conformal Killing fields which vanish at $p_j$
but which do not vanish identically on the component of 
$\Sigma$ containing $p_j$, for $j=1,2$.
\end{itemize}
\end{definition}

\begin{remark} To avoid trivialities, we always assume that either $\Sigma$
is connected or else has two components, each of which contains one of the 
points $p_j$. Note that this nondegeneracy condition is very mild and holds 
for any fixed conformal class $[\gamma]$ when the points $p_j$ are chosen
generically. In the asymptotically Euclidean and asymptotically
hyperboloidal cases, which are the only noncompact  
cases of interest to us here, 
this nondegeneracy condition is not needed.  We discuss these cases in 
more detail in \S 7.
\end{remark}

Assume that $\Sigma$ is compact, and fix two points $p_1$, $p_2$ on it.
Suppose that $(\gamma,\Pi)$ is an initial data set with $\Pi\not\equiv
0$ and $\tau=\tr_\gamma \Pi$ constant, so that in particular the two
constraint equations (\ref{eqn1}), (\ref{eqn2}) are satisfied. It is
often more convenient to regard $(\gamma,\mu)$ as the initial data
set instead, where $\mu$ is the transverse-traceless part of $\Pi$. For any 
(small) $R>0$, fix balls $B_j=B_R(p_j)$ of radius $R$ 
(with respect to $\gamma$) around the points $p_j$, and define
\begin{equation}
\Sigma_R^* = \Sigma \setminus (B_1 \cup B_2).
\label{eq:defsgr}
\end{equation}
We also let $\Sigma^*_0 = \Sigma^* = \Sigma \setminus \{p_1,p_2\}$. 
We shall modify the metric $\gamma$ conformally in each of the
punctured balls $B_j \setminus \{p_j\}$ to obtain a metric $\gamma_c$ on
$\Sigma^*$ with two asymptotically cylindrical ends. These ends have a 
natural parameter $t = -\log r$, where $r$ is the geodesic distance in $B_j$ 
to $p_j$. According to (\ref{eq:conf2}) the transverse-traceless tensor 
$\mu$ changes to a tensor $\mu_c$ which is transverse-traceless with
respect to $\gc$. This gives an asymptotically cylindrical
pair $(\gamma_c,\mu_c)$ together with a function $\psi$ on $\Sigma^*$
which solves the associated Lichnerowicz equation (\ref{eqn3}) (with
coefficients determined by this pair of tensors). We also refer to the
triple $(\gamma_c,\mu_c,\psi)$ as an initial data set.

The construction proceeds as follows. We first identify long
pieces, of length $T$, of these two cylindrical ends with one
another to obtain a family of metrics $\gamma_T$ on the 
new manifold $\ST$ obtained from $\Sigma$ by
adding a handle. For each $T$, $\ST$ is of course diffeomorphic to 
the manifold $\widehat{\Sigma}$ described in the previous section. 
This process involves cutoff functions and is not canonical. We also use 
cutoffs to patch together the values of $\mu_c$ on these ends to obtain
$\mu_T$, and similarly an approximate solution $\psi_T$ of the
Lichnerowicz equation. We write this out more carefully in \S \ref{se:2}. 
Although it is not hard to construct $\mu_T$ so that it still has 
vanishing trace relative to $\gamma_T$, it is no longer necessarily 
true that $\mu_T$ will be divergence-free. 
The triple $(\gamma_T,\mu_T,\psi_T)$ does not solve the constraint 
equations, but instead gives an approximate solution, with
an error term tending to zero as $T \to \infty$.  We show that
it is possible, for $T$ sufficiently large, to correct this data
to obtain exact solutions of the constraint equations. We do this in
the following order. First $\mu_T$ is altered by a small correction
term $\sigma_T$, using a well known second order linear elliptic system 
(usually called the vector Laplacian) 
derived from the conformal Killing operator on 
$(\ST,\gamma_T)$, to obtain a tensor ${\tilde\mu}_T = 
\mu_T-\sigma_T$ which is transverse-traceless with respect to 
$\gamma_T$. This involves a careful analysis 
of the mapping properties of this operator with good control
as $T \to \infty$ and is the topic of \S \ref{se:3}. We then use the conformal
method, and more specifically the Lichnerowicz equation (\ref{eqn3}), 
to modify the function $\psi_T$ to some $\tpt$, so that 
$(\gamma_T,{\tilde \mu}_T,\tpt)$ is an initial data set, i.e. gives
an exact solution of the constraint equations. This involves first 
estimating the error 
terms for this approximate solution, which is done in \S \ref{se:4}.
We next analyze the linearization of the Lichnerowicz equation
uniformly as $T \to \infty$. In particular, we introduce a class of
weighted H\"older spaces ${\mathcal C}^{k,\alpha}_{\delta}(\ST)$ which 
correspond to the standard H\"older spaces away from the handle and 
determine the range of weights $\delta$ for which the inverse of the
linearized Lichnerowicz equation is uniformally bounded independent of
$T$.  This is carried out in \S \ref{se:5}.  Finally, in \S \ref{se:6}
we use a contraction 
mapping argument to produce the exact solutions by finding a
function $\eta_T$ in the  ball ${\mathbb B}_\nu$ of radius
$\nu\,e^{-T/4}$ in ${\mathcal C}^{k+2,\alpha}_{\delta}(\ST)$ for
$\delta\in (0,1)$, so that $\tpt=\pt+\eta_T$ satisfies the Lichnerowicz
equation. This establishes our main theorem which we now state
(in the form appropriate for initial data on closed manifolds).
\begin{theorem}
Let  $(\Sigma, \gamma, \Pi; p_1, p_2)$ be a compact, smooth, marked, 
constant mean curvature solution of the Einstein constraint 
equations; thus $\Pi=\mu+\frac{1}{3}\tau\gamma$ with $\tau$ constant 
and $\mu$ transverse-traceless with respect to $\gamma$. We assume that 
this solution is nondegenerate and also that $\Pi\not\equiv 0$. 
Then there is a geometrically natural choice of a parameter $T$ and, for
$T$ sufficiently large, 
a one-parameter family of solutions $(\ST, \Gt, \Pi_{T})$ of the Einstein 
constraint equations with the following properties. First, the 
three-manifold $\Sigma_T$ is constructed from $\Sigma$ by adding
a handle, or neck, connecting the two points $p_1$ and $p_2$. Next, 
$$
\Gt= \tpt^{4}\,\gt \qquad \mbox{ and }\qquad 
\Pi_{T} =  \tpt^{-2}\,\tmut +\frac{1}{3}\tau\tpt^{4}\gt,
$$ 
where $\tpt=\pt+\eta_{T}$ and $\tilde{\mu}_T = \mu- \sigma_T$
are very small perturbations of the conformal factor $\pt$ and
approximate transverse-traceless tensor $\mu_T$ defined informally 
above and more carefully in \S 2 below. The tensor $\tilde{\mu}_T$
is transverse-traceless with respect to $\Gamma_T$ and the 
function $\tpt$ satisfies the Lichnerowicz equation with respect to 
the pair $(\gt,\tmut)$. Finally, the perturbation 
terms satisfy the following estimates as $T \to \infty$: 
\begin{itemize}
\item On $\Sigma^{*}_{R}$, 
\[
|\Gamma_T - \gamma|_{\gamma} \leq Ce^{-T/4}, \qquad
|\Pi_T - \Pi|_{\gamma} \leq C T^3 e^{-3T/2};
\]
\item On the neck region, there is a natural choice of coordinate
system $(s,\theta)$, $-T/2 \leq s \leq T/2$ and $\theta \in S^2$,
such that if $h_0$ is the standard metric on $S^2$, then
\[
\gamma_T = ds^2 + h_0 + {\mathcal O}(e^{-T/2}\cosh s),
\]
\[
\psi_T = 2e^{-T/4}\cosh(s/2) + {\mathcal O}(e^{-T/4}),
\qquad |\eta_T| \leq Ce^{-(1+\delta) T/4}(\cosh(s/2))^{\delta}
\]
for some weight $\delta \in (0,1)$, and finally 
\[
|\tilde{\mu}_T|_{\gamma_T} \leq C e^{-T/2}\cosh s,
\qquad |\sigma_T|_{\gamma_T} \leq CT^3 e^{-3T/2}.
\]
\end{itemize}
\label{thm:1}
\end{theorem}
The modifications needed for the asymptotically Euclidean
and asymptotically hyperboloidal cases are discussed in \S \ref{se:7} 
and statements of this theorem in these settings are given there,
in Theorems \ref{th:AE} and \ref{th:AH}. 

Notice that we have not stated the precise regularity of the solutions, 
but as indicated earlier, we work in various classes of H\"older spaces; 
in particular, we also obtain existence of initial data sets
$(\Sigma, \gamma, \Pi)$ with only finite regularity.

Some estimates of the geometry of the solutions are provided in \S 8;
these should be of interest when considering the spacetime evolutions
of these initial data sets. Finally, \S 9 contains an informal
discussion of various concrete examples of initial data sets, many 
of which are new, which our construction gives. 

\section{The approximate solution} 
\label{se:2}

In this section we construct the family of approximate solutions 
$(\gamma_T,\mu_T,\psi_T)$. Let $r_j$ denote the geodesic distance 
to $p_j$ (relative to $\gamma$) in the ball $B_j$. By Gauss' lemma 
we have 
\[
\left. \gamma \right|_{B_j} = dr_j^2 + r_j^2 h_j(r_j),
\]
where $h_j$ is a family of metrics on $S^2$, smooth down to 
$r_j=0$ and such that $h_j(0)$ is the standard round metric on 
the sphere. The standard way to obtain a metric with asymptotically 
cylindrical ends is to multiply $\gamma$ by $r_j^{-2}$ in each
of these balls. Recalling the convention of (\ref{eq:conf1}) that
the conformal factor is written as $\psi^4$, let us fix a function
$\psi_c$ which is strictly positive on $\Sigma^*$, which equals
one on $\Sigma^*_{2R}$, and which equals $r_j^{1/2}$ on $B_j$.
Now define 
\begin{equation}
\gamma_c = \psi_c^{-4}\gamma.
\label{eq:3.2}
\end{equation}
This metric agrees with $\gamma$ on $\Sigma^*_{2R}$ and has
asymptotically cylindrical ends in a neighborhood of each $p_j$. 

\medskip

\def\pa{$\scriptstyle{\partial B_{2R}(P_1)}$}
\def\pb{$\scriptstyle{\partial B_R(P_1)}$}
\def\pc{$\scriptstyle{P_1}$}
\def\paa{$\scriptstyle{\partial B_{2R}(P_2)}$}
\def\pbb{$\scriptstyle{\partial B_R(P_2)}$}
\def\pcc{$\scriptstyle{P_2}$}
\def\pe{$\scriptstyle{t_1=A}$}
\def\pf{$\scriptstyle{t_1=A+\frac{T}{2}}$}
\def\pg{$\scriptstyle{t_1=A+T}$}
\def\pgg{$\scriptstyle{t_2=A+T}$}
\def\pff{$\scriptstyle{t_2=A+\frac{T}{2}}$}
\def\pee{$\scriptstyle{t_2=A}$}

\begin{center}
\begin{psfrags}
\includegraphics[height=5truein]
{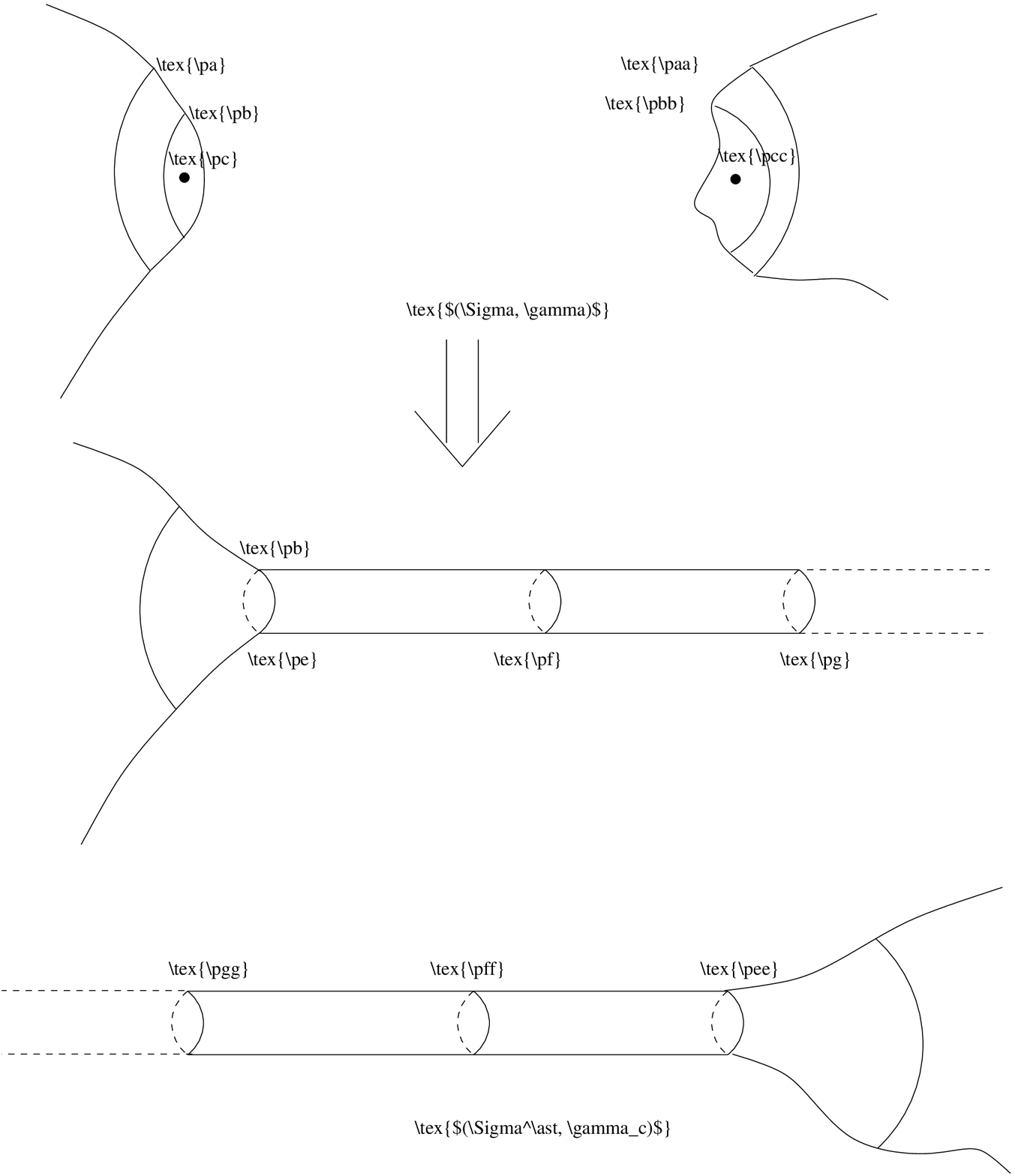}
\end{psfrags}\\[3mm]
{\bf Figure 1}: Conformally blowing up $(\Sigma,\gamma)$ to form
$(\Sigma^\ast, \gamma_c)$ with two asymptotically cylindrical ends.
\end{center}

In fact, setting $t_j = -\log r_j$, we see that
\[
\left. \gamma_c \right|_{B_j} = dt_j^2 + h_j(e^{-t_j}).
\]
Recalling that $h_j$ is smooth in $r_j$, we can rewrite this as
\begin{equation}
dt_j^2 + d\theta^2 + e^{-t_j}k_j,
\label{eq:mcyl}
\end{equation}
where $d\theta^2$ is the standard round metric on $S^2$ and
the term $k_j$ in the remainder is a symmetric two-tensor 
which is smooth in $r_j=e^{-t_j}$.
Note that if the metric $\g$ is conformally flat in $B_j$ then the
definition of $\psi_c$ may be easily modified to make the metric $\gc$
exactly cylindrical in $B_j$.
Note also, for later reference, that 
\begin{equation}
-\log R \leq t_j < \infty.
\label{eq:rtj}
\end{equation}

Next, following (\ref{eq:conf2}), define 
\begin{equation}
\mu_c = \psi_c^{2}\mu;
\label{muic-size} 
\end{equation}
this is defined on $\Sigma^*$ and is transverse-traceless with respect 
to $\gamma_c$. Relative to the initial data set $(\gamma_c,\mu_c)$
on $\Sigma^*$, $\psi_c$ satisfies the Lichnerowicz equation
\begin{equation}
\Delta_c \psi_c - \frac{1}{8}\, R_c \,\psi_c +\frac{1}{8}\,
|\mu_c|^2\, \psi_c^{-7} - \frac{1}{12}\, \tau^2\,\psi_c^5 = 0.
\label{eq:3.3}
\end{equation}
In this equation, the expressions $\Delta_c$, $R_c$ and $|\cdot|^2$ 
are all computed in terms of the metric $\gamma_c$. Thus
$(\gamma_c,\mu_c,\psi_c)$ is an initial data set on $\Sigma^*$. 

Let $A=-\log R$ and let $T$ be a large
parameter. Truncate the manifold $\Sigma^*$ by omitting the regions
where $t_j > A + T$. We may form a new smooth manifold by
identifying the two finite cylindrical tubes $\{(t_j,\theta): A \leq t_j 
\leq A+T\}$ via the map $(t_1,\theta) \to (T-t_1,-\theta)$.
Let us call the resulting manifold $\ST$ and
let the now-identified cylindrical tube be denoted $C_T$. (Of course,
each $\ST$ is diffeomorphic to $\widehat{\Sigma}$, the manifold
obtained by adding a handle to $\Sigma$, but it is convenient to
keep track of the dependence of $T$ explicitly.) We introduce
a new linear coordinate $s$ on $C_T$ by setting 
\[
s = t_1 - A - T/2 = -t_2 + A + T/2.
\]
Thus $C_T$ is parametrized by the coordinates $(s,\theta)
\in [-T/2,T/2] \times S^2$. 

\medskip

\def\st{$\scriptstyle{s=-\frac{T}{2}}$}
\def\so{$\scriptstyle{s=0}$}
\def\stt{$\scriptstyle{s=\frac{T}{2}}$}

\begin{center}
\begin{psfrags}
\includegraphics[height=2truein]
{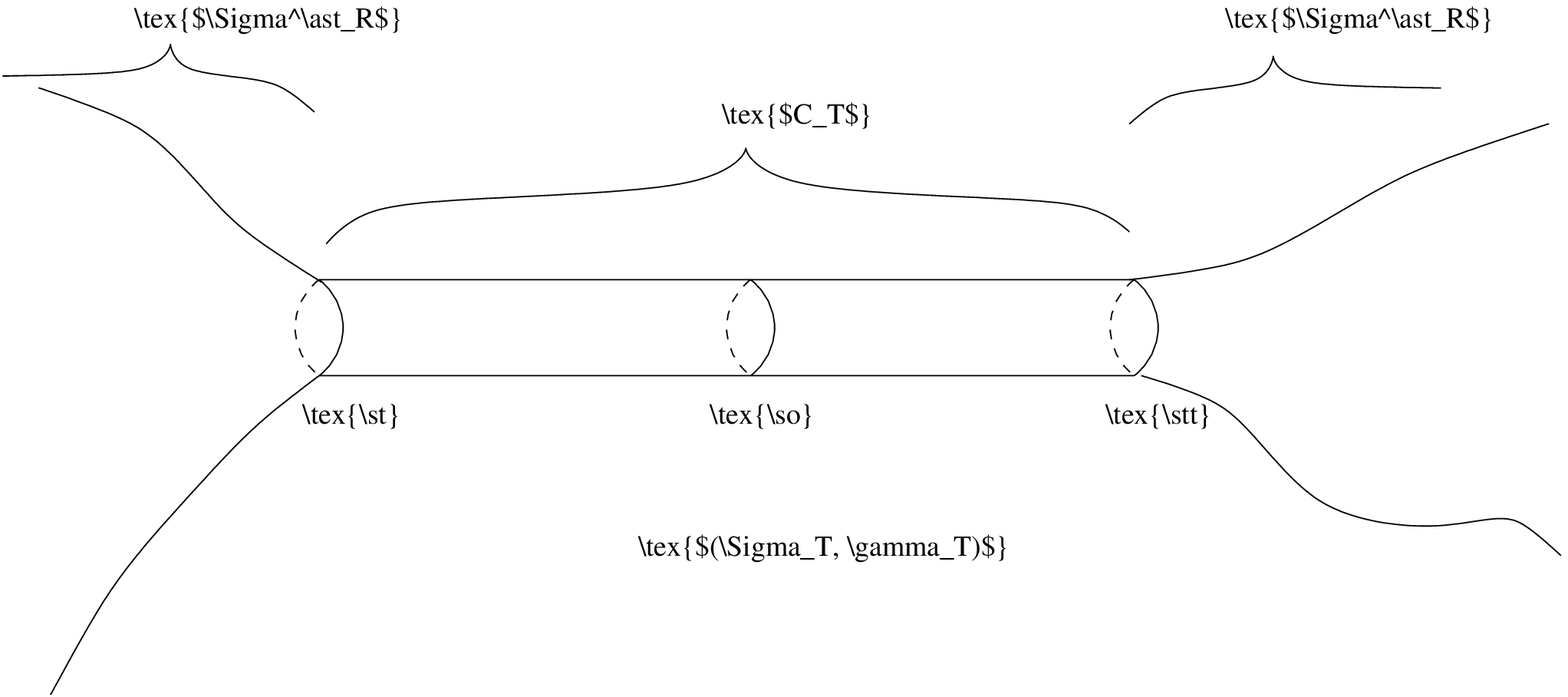}
\end{psfrags}\\[3mm]
{\bf Figure 2}: The manifold $(\Sigma_T, \gamma_T)$.
\end{center}

To continue we must define the family of metrics $\gamma_T$, 
`almost-transverse-traceless' tensors $\mu_T$ on $\ST$,
and conformal factors $\psi_T$, which we do in turn.

The definition of $\gamma_T$ is most transparent when $\gamma$ is 
conformally flat in the balls $B_j$. In this case, using the
modifications of $\psi_c$ alluded to above, in the cylindrical 
coordinates $(t_j,\theta)$, we have $\gamma_c = dt_j^2 + d\theta^2$, 
and thus the map identifying 
the two cylindrical pieces to $C_T$ is an isometry and $\gamma_T$
is thereby well-defined. In the general case, we cover $\ST$ by two open
sets ${\mathcal U}_1$ and ${\mathcal U}_2$, the intersection of which
consists of two components, one disjoint from $B_1 \cup B_2$ and 
the other equal to $\{(s,\theta): -1 < s < 1\}$. Choose a partition 
of unity $\{\chi_1,\chi_2\}$ subordinate to this cover and let 
$\gamma_j$ and $\mu_j$ denote the restrictions of $\gamma_c$ and $\mu_c$
to ${\mathcal U}_j$. Then, with the obvious abuse of notation, define 
\[
\gamma_T = \chi_1 \gamma_1 + \chi_2 \gamma_2,
\qquad
\mu_T = \chi_1 \mu_1 + \chi_2 \mu_2.
\]
Notice that $\gamma_T = \gamma_c$ and $\mu_T = \mu_c$ away from the middle
$Q = [-1,1] \times S^2$ of the neck region, where $t_1, t_2 \approx T/2$.

\medskip

\begin{center}
\begin{psfrags}
\includegraphics[height=2truein]
{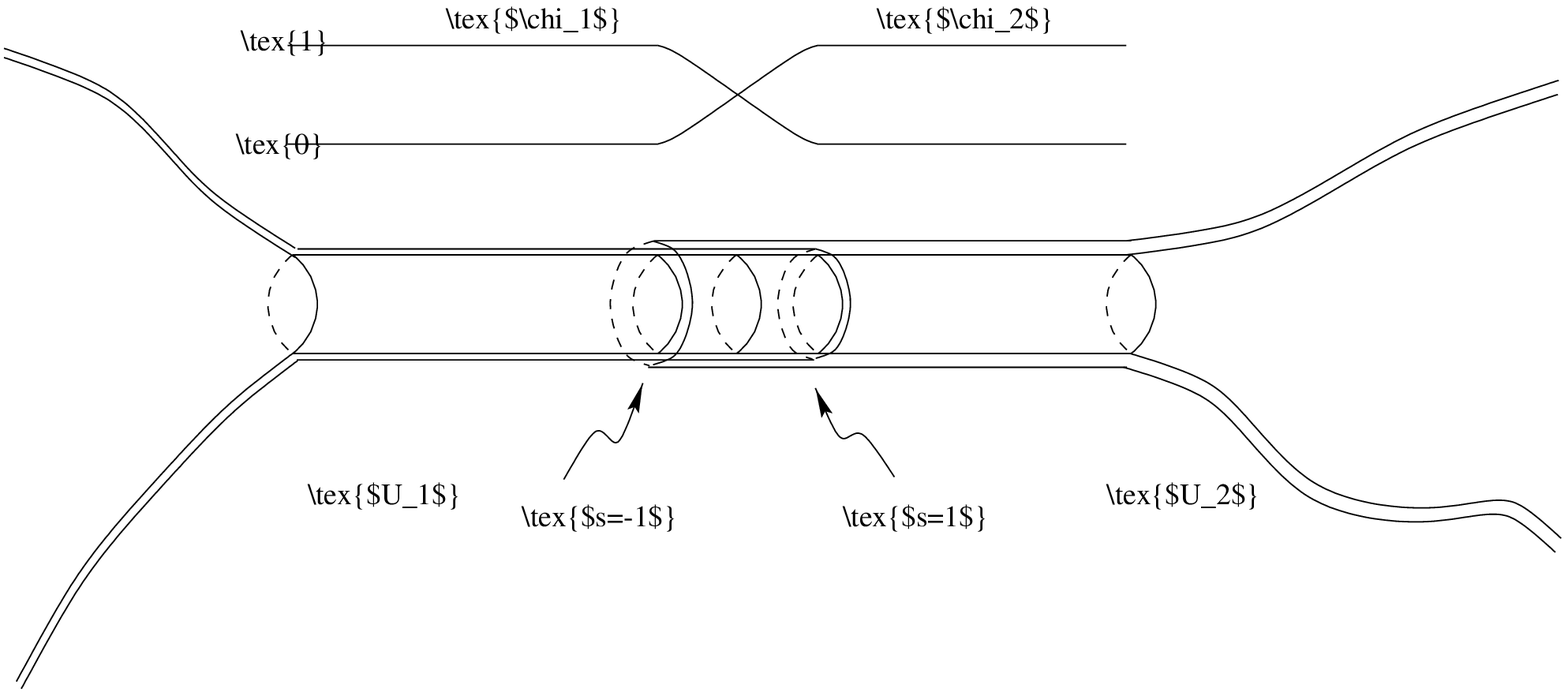}
\end{psfrags}\\[3mm]
{\bf Figure 3}: Defining the metric 
$\gamma_T = \chi_1 \gamma_1 + \chi_2 \gamma_2$,
and the approximate TT-tensor $\mut = \chi_1 \mu_1 + \chi_2
\mu_2$.
\end{center}

Although it is not obvious why at this stage, we need to define $\psi_T$ 
somewhat differently. This time choose a covering 
$\{\tilde{\mathcal U}_1,\tilde{\mathcal U}_2\}$ of $\Sigma$
such that $\tilde{\mathcal U}_1 \cap \tilde{\mathcal U}_2
\subset \Sigma_{2R}^*$ and with $p_j \in \tilde{\mathcal U}_j$.
Also choose nonnegative functions $\tilde{\chi}_j \in 
{\mathcal C}^\infty_0(\tilde{\mathcal U}_j)$ with $\tilde{\chi}_1 + 
\tilde{\chi}_2 = 1$ on $\Sigma_R^*$, and such that when restricted to $B_j$, 
$\tilde{\chi}_j = 1$ for $t_j \leq A+T-1$ and $\tilde{\chi}_j = 0$ for 
$t_j \geq A+T$. Writing the restriction of $\psi_c$ to $\tilde{\mathcal U}_j$ 
as $\psi_j$, we set 
\[ 
\psi_T = \tilde{\chi}_1 \psi_1 + \tilde{\chi}_2 \psi_2.
\] 
Then, based on the construction of $\ST$ from the conformal blow up of
$(\Sigma, \gamma)$, we define $\pt$ on $\ST$ in the obvious way. Note
that $\pt$ is exactly equal to $1$ away from the cylinder, and is equal to
$\psi_1+\psi_2$ throughout most of the cylinder except at the two ends.
At these ends, near the junctions of the cylinder with $\SR$, one of 
the two functions $\psi_1$ or $\psi_2$ is roughly of unit size and the 
other is exponentially small; in defining $\pt$, the exponentially small
summand is cut off to be zero near these junctures.

\medskip

\begin{center}
\begin{psfrags}
\includegraphics[height=2truein]
{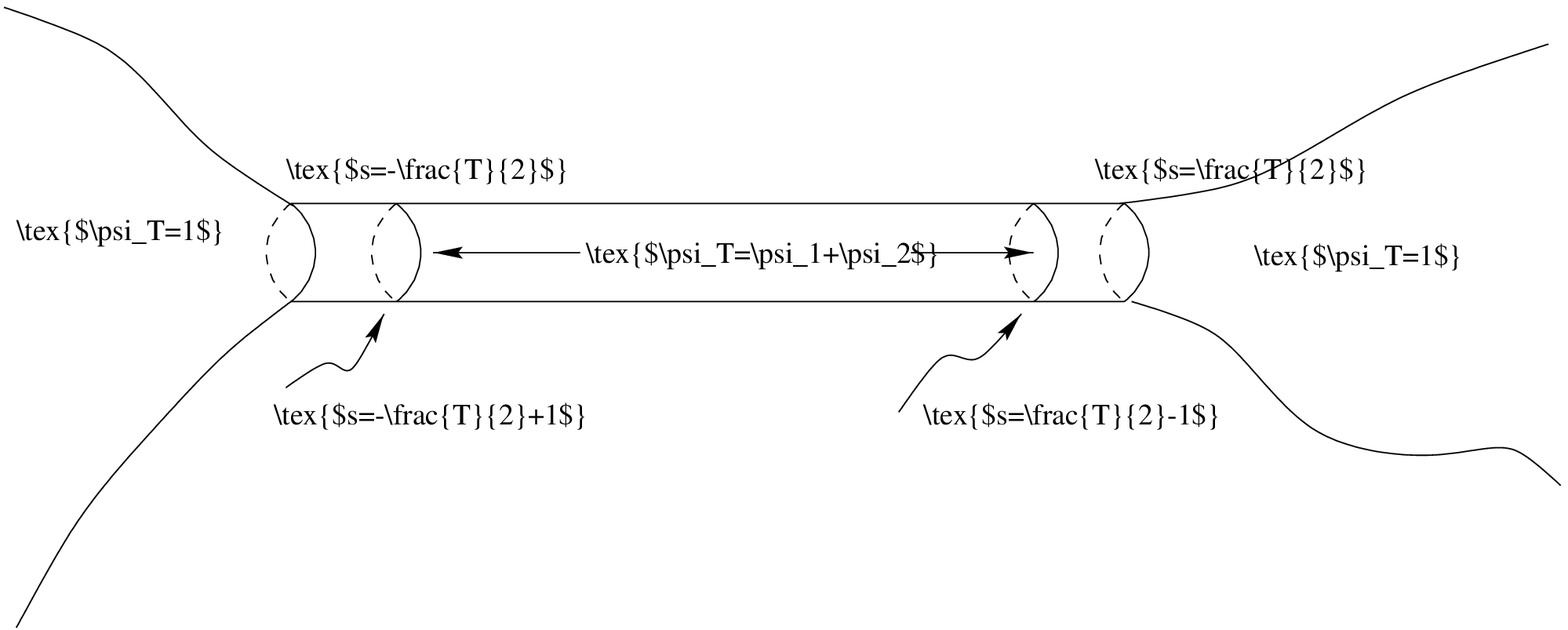}
\end{psfrags}\\[3mm]
{\bf Figure 4}: The approximate solution to the Lichnerowicz equation,
$\psi_T=\widetilde \chi_1\psi_1+\widetilde \chi_2\psi_2$.
\end{center}

In summary, we have defined $\gamma_T$ and $\mu_T$ by
using cutoffs supported on the middle of the cylinder $C_T$,
but have cut off either of the $\psi_j$ at the ends of
the cylinder. The reasons for doing this are that $\gamma_c$
decays to the product metric like $e^{-t_j}$ and $\mu_c$ is
extremely small in the middle of $C_T$, so the error terms
from these are of order $e^{-T/2}$. On the other hand,
$\psi_c$ only decays like $e^{-t_j/2}$ along each end, and
so if we were to cut this function off in the middle, it
would produce an error term with size approximately $e^{-T/4}$,
which is unacceptably large. We carry out these estimates in more
detail in \S \ref{se:4}.

\section{Transverse-traceless tensors on $\ST$}
\label{se:3}

We now undertake the first of several steps intended to perturb the 
approximate solution $(\gamma_T,\mu_T,\psi_T)$ constructed in the 
last section to an initial data set, i.e. an exact solution, when $T$ 
is sufficiently large. Our goal in this section is to modify the 
tensor $\mu_T$ so that it becomes transverse-traceless with respect to 
$\gamma_T$. 

Before starting, note that it is clear from the definitions that $\mu_T$ 
is transverse-traceless away from the center of the neck region $Q$ in the 
middle of the cylinder $C_T$. Furthermore, by choosing the cutoff
functions carefully, we may also assume that $\mu_T$ is trace-free
on all of $\Sigma_T$; however, its divergence is almost surely
nonzero, but at least is supported in the annulus $Q$. 

There is a well-known procedure for producing a correction term to kill 
this divergence. We recall this now, and refer to \cite{CBY} or \cite{T3}
for more details. Let $\gamma$ be a metric and $X$ a vector 
field on the three-manifold $\Sigma$, and write
\[
\D X = \frac{1}{2}{\mathcal L}_X \gamma - \frac{1}{3}(\di X)\gamma.
\]
Equivalently, in local coordinates,
\[
(\D X)_{jk}=\frac{1}{2}\left(X_{j;k}+X_{k;j}\right)-\frac{1}{3}\dive (X)\,
\gamma_{jk}.
\]
The first order operator $\D = \D_\gamma$ defined by this expression
maps vector fields to trace-free symmetric $(0,2)$ tensors and has the
property that $\D X = 0$ if and only if $X$ is a conformal Killing field.
The formal adjoint of $\D$ is $\D^{*} = -\dive$, and the second order 
operator $L = \D^{*}\circ\D$ is self-adjoint, nonnegative and elliptic. 
This operator is often referred to as the `vector Laplacian'.

Now let $\gamma = \gamma_T$, with $\D$ and $L$ the associated operators. 
Let $W$ be the vector field associated to $\dive \mu_T$ using the natural 
duality between vector fields and $1$-forms. Suppose that we can solve the 
equation $LX = W$. Then writing $\sigma_T = \D X$ we compute that
\[
\dive (\mu_T - \sigma_T) = W - \dive \D X = W - L X = 0,
\]
and so $\mu_T - \sigma_T$ is divergence-free. Since $\mu_T$ and $\sigma_T$ 
are both trace-free, we have produced the desired correction term.

To implement this strategy properly, we must show that when $T$
is large enough, the operator $L$ is invertible, and moreover
that the solution $X$, and hence $\D X$, is much smaller than $\mu_T$,
so that $\sigma_T$ can honestly be regarded as a small perturbation.
Note that since $W$ is a divergence and hence orthogonal to the
cokernel of $L$, we can always solve $LX=W$ (but without an estimate on the
solution).
The main issue is to prove that the inverse $G$ to $L$ on the
manifold $\Sigma_T$, when acting between appropriate function spaces,
has norm bounded at worst by some polynomial in $T$. (This is
sufficient because the error term $\dive \mu_T$ decays exponentially.)

When $\gamma$ is conformally flat in the balls $B_j$, it is possible
to choose the conformal factor $\psi$ used to define $\gamma_c$ slightly 
differently so that $\gamma_T$ is the product metric on the cylinder 
$C_T = [-T/2,T/2] \times S^2$. We proceed in this case by an analysis of 
the operators $\D$ and $L$ on the complete cylinder $\RR \times S^2$ as 
well as on the finite cylindrical piece $C_T$, using separation of 
variables and explicit calculations with the resulting ODEs, to reach the 
conclusion that the inverse $G$ to $L$ exists and has norm (as an operator 
on $L^2(\Sigma_T)$) blowing up no faster than $T^2$. When $\gamma$ is
not conformally flat in these balls, an additional perturbation argument
is required to reach the same conclusion. The basic goal in these arguments
is to show that the lowest Dirichlet eigenvalue of $L$ on $C_T$ is of order 
$T^{-2}$. Afterwards, the same conclusion is reached for the operator 
$L$ on $\Sigma_T$. 

\subsection{The conformal Killing operator on the cylinder} 
\label{L-cyl}

\subsubsection{The decomposition}
We first derive the explicit form of $L$ on the cylinder $\RR \times S^2$
with product metric $ds^2 + h$, where $s$ is the linear coordinate on $\RR$
and $h$ is the standard round metric on $S^2$. Any vector field $X$ on the 
cylinder may be written 
$$
X = f\, \partial_{s} + Y(s)
$$
where $f$ is a function on the cylinder and $Y(s)$ is a vector field on the 
sphere $\{s\}\times S^2$. Write $\D X = S(X) - \frac{1}{3}\di X\,\g$, and 
choose any local coordinates $\theta = (\theta_1,\theta_2)$ on $\Sp^2$. Then 
for $i,j = 1,2$, 
$$
S(f\partial_s)_{00} =\partial_s f,\qquad 
S(f\partial_s)_{0i} = \frac{1}{2} f_{i}, \qquad
S(f\partial_s)_{ij}=0
$$ 
and $\di(f\partial_s)=\partial_s f$. Therefore, exhibited as a matrix, 
\[
\D(f\partial_s)= 
\left(
\begin{array}{cc}
\frac{2}{3}\partial_s f & \frac{1}{2}\nabla_{\theta} f  \\
 &   \\
\frac{1}{2}\nabla_{\theta} f  
& -\frac{1}{3}(\partial_s f) \, h 
\end{array}
\right),
\]
where $\nabla_\theta$ represents the gradient of $f$ on the sphere.
Hence, with the convention $\Delta_\theta = -\di\,\nabla_\theta$, 
\[
L(f\partial_s)  =  -\di(\D(f\partial_s)) = \left(
\begin{array}{c}
-\frac{2}{3}\partial^2_s f
 +\frac{1}{2}\Delta_{\theta}f \\
\\
-\frac{1}{2}\partial_s (\nabla_{\theta} f) + \frac{1}{3} 
\nabla_{\theta}(\partial_s f) 
\end{array}
\right) 
=  
\left(
\begin{array}{c}
-\frac{2}{3}\partial^2_s f
 +\frac{1}{2}\Delta_{\theta}f \\
\\
-\frac{1}{6}\partial_s (\nabla_{\theta} f) 
\end{array}
\right).
\]
Next,
$$
S(Y)_{00} = 0, \qquad
S(Y)_{0i} = \frac{1}{2}\partial_s(Y_i), \qquad
S(Y)_{ij} = S_{\theta}(Y)_{ij},
$$
where by definition the conformal Killing operator $\D_h$ on the sphere
decomposes as $\D_h Y = S_\theta(Y) - \frac{1}{2} (\di_\theta Y) h$.
Using also that $\di(Y) = \di_{\theta}(Y)$, we get
\[
\D(Y)= 
\left(
\begin{array}{cc}
-\frac{1}{3}\di_{\theta}(Y) & \frac{1}{2}\partial_s Y   \\
 &  \\
\frac{1}{2}\partial_s Y 
& \D_{\theta}(Y) +\frac{1}{6}\di_{\theta}(Y)\, h 
\end{array}
\right),
\]
and then that
\[
L(Y)  =  -\di(\D(Y)) =  
\left(
\begin{array}{c}
\frac{1}{3}\partial_s (\di_{\theta}(Y))
 -\frac{1}{2}\di_{\theta}(\partial_s Y)\\
\\
-\frac{1}{2} \partial^2_s Y -\di_{\theta}(\D_{\theta}(Y))
-\frac{1}{6} \nabla_{\theta}\di_{\theta}(Y)
\end{array}
\right)
\]
\[
=  
\left(
\begin{array}{c}
-\frac{1}{6} \partial_s (\di_{\theta}(Y))\\
\\
-\frac{1}{2} \partial^2_s Y + L_{\theta}(Y) - \frac{1}{6}
\nabla_{\theta}\di_{\theta}(Y)
\end{array}
\right).
\]

It is possible to express this operator in a somewhat more familiar 
form. To do this, we identify vector fields and $1$-forms on $S^2$
in the usual way using the metric $h$. Thus, $\nabla_\theta$ and $\di_\theta$ 
correspond to $d_\theta$ and $-\delta_\theta$, respectively. We still let $Y$ 
denote the $1$-form associated to the vector field $Y$, hopefully without 
causing undue confusion. Using these identifications,
\[
L(f\, ds + Y(s)) =  
\left(
\begin{array}{c}
-\frac{2}{3}\partial^2_s f 
 + \frac{1}{2}\Delta_{\theta}f  +\frac{1}{6}\partial_s(\delta_{\theta}Y)\\
\\
-\frac{1}{2} \partial^2_s Y  +L_{\theta}(Y) +
\frac{1}{6} d_{\theta}\delta_{\theta}(Y) -
\frac{1}{6} \partial_s (d_{\theta} f)
\end{array}
\right).
\]
This expression may be simplified using two separate Weitzenb\"ock formul\ae, 
\[
L_{\theta} = \frac{1}{2} \nabla^{*}\nabla -\frac{1}{2},
\qquad \mbox{and}\qquad 
\Delta_\theta = \nabla^{*}\nabla  + 1,
\]
both of which use the fact that the curvature tensor on $S^2$ is constant. 
The first of these relates $L_\theta$ to the `rough Laplacian' 
$\nabla_\theta^*\nabla_\theta$ (cf. \cite{T3}) and the second is the more 
familiar one relating this rough Laplacian to the Hodge Laplacian 
$\Delta_\theta = d_\theta \delta_\theta + \delta_\theta d_\theta$. These 
combine to give 
\[
L_{\theta} = \frac{1}{2}\Delta_\theta -1,
\]
and thus we finally arrive at the expression
\begin{eqnarray}
L\left(\begin{array}{c}
f\\  \\ Y
\end{array}
\right)&=& 
\left(
\begin{array}{cc}
-\frac{2}{3}\partial^2_s  + \frac{1}{2}\Delta_{\theta}  &
\frac{1}{6}\partial_s\,\delta_{\theta}    \\
  &   \\
-\frac{1}{6}\partial_s\, d_{\theta} &  
 -\frac{1}{2}\partial^2_s + \frac{2}{3}d_{\theta}\, \delta_{\theta}
+\frac{1}{2}\delta_{\theta}\, d_{\theta} -1 
\end{array}
\right)\left(\begin{array}{c} f\\  \\ Y \end{array}\right).\label{L-op}
\end{eqnarray}

\subsubsection{Separation of variables}
To analyze this operator $L$ on the cylinder in more detail, we introduce the
eigenfunction expansion on the $S^2$ factor. Let $\{\phi_k\}$ be an 
orthonormal set of eigenfunctions for the scalar Laplacian on $S^2$, so that
$\Delta_\theta \phi_k = \lambda_k \phi_k$ and each $\lambda_k=j(j+1)$ for some
$j \in {\mathbb N}$. If $*$ is the Hodge star on $S^2$, then $\{\lambda_k^{
-1/2}d_\theta \phi_k, \lambda_k^{-1/2}\delta_\theta *\phi_k\}$ is an 
orthonormal basis of eigenfunctions for the Hodge Laplacian on $1$-forms. Set 
$\phi=\phi_k$ and $\lambda = \lambda_k$ with $\lambda_k > 0$, for some $k$, 
and consider a $1$-form  of the form 
\[
X =  f(s,\theta)ds + Y(s) =  u(s)\phi\,ds + 
v(s)\frac{d_{\theta}\phi}{\sqrt{\lambda}} + 
w(s)\frac{\delta_{\theta}*\phi}{\sqrt{\lambda}},
\]
or equivalently, the associated dual vector field. Then 
\[
LX = \left(-\frac{2}{3} u^{\prime\prime} + \frac{\lambda}{2} u + 
\frac{\sqrt{\lambda}}{6} v^{\prime}\right)\phi\,ds + 
\left(-\frac{1}{2} v^{\prime\prime} + (\frac{2}{3}\lambda -1)v 
-\frac{\sqrt{\lambda}}{6} u^{\prime}\right) \frac{d_{\theta}\phi}{\sqrt{
\lambda}} 
\]
\[
+ \left(-\frac{1}{2} w^{\prime\prime} + (\frac{1}{2}\lambda -1)w
\right) \frac{\delta_{\theta}*\phi}{\sqrt{\lambda}},
\]
and so $L$ acts on the column vector $(u,v,w)^t$ by
\[
\left(\begin{array}{ccc}
-\frac{2}{3}  & 0 & 0 \\
0 & -\frac{1}{2} & 0 \\
0 & 0 & -\frac{1}{2}\end{array}\right)\del_s^2 + 
\left(\begin{array}{ccc}
0 & \frac{\sqrt{\lambda}}{6} & 0 \\
-\frac{\sqrt{\lambda}}{6} & 0 & 0 \\
0 & 0 & 0 \end{array} \right)\del_s + 
\left(\begin{array}{ccc}
\frac{\lambda}{2} & 0 & 0 \\
0 & \frac{2}{3}\lambda - 1 & 0 \\
0 & 0 & \frac{1}{2}\lambda - 1 \end{array} \right).
\]
There is one extra case, when $\lambda = \lambda_0 = 0$; then $X = u(s) 
\,ds$ and $L X = -\frac{2}{3}u''(s)ds$, so we write
\[
L_0 = -\frac{2}{3}\del_s^2.
\] 

When $j\geq 1$, $L_j$ uncouples as $L_j = L'_j \oplus L''_j$ where
\[
L_j' = \left(\begin{array}{cc} -\frac{2}{3}  & 0 \\
0 & -\frac{1}{2} \end{array}\right)\del_s^2 + 
\left(\begin{array}{cc}
0 & \frac{\sqrt{\lambda}}{6} \\
-\frac{\sqrt{\lambda}}{6} & 0 \end{array} \right)\del_s + 
\left(\begin{array}{cc}
\frac{\lambda}{2} & 0 \\ 0 & \frac{2}{3}\lambda - 1 
\end{array} \right)
\]
acts on $(u,v)^t$ and
\[
L_j'' = -\frac12 \del_s^2 + (\frac12 \lambda - 1)
\]
acts on $w$. Thus altogether,
\[
L = L_0 \oplus \bigoplus_{j \geq 1} (L'_j \oplus L''_j).
\]

\subsubsection{Temperate solutions}
The first application of these formul\ae\ concerns the solutions of
$LX=0$ on the cylinder which do not grow exponentially in either 
direction.

\begin{proposition}
The space of temperate solutions to $LX=0$ on the cylinder $\RR \times S^2$ 
is 8-dimensional and is spanned by the 2-dimensional family of 
$\theta$-independent radial vector fields $(a+bs)\partial_s$, $a,b\in\R$, and 
the 6-dimensional family of vector fields which are dual to the $1$-forms 
$(c_i+d_is)\delta_{\theta}*\phi_i$, $c_i,d_i\in\R$, where $\phi_i=x_i|_{S^2}$, 
$i=1,2,3$.
\end{proposition}
{\bf Proof:} It is straightforward to check that these bounded or linearly 
growing vector fields are all in the nullspace of $L$. In fact, using the
eigendecomposition of $L$ above, $L_0 u = -\frac{2}{3}u''= 0$ implies 
$u(s) = a + bs$, and this gives the first family. Next, for $j=1$ (i.e. 
$\lambda=2$), $L''_1 w = -\frac{1}{2}w'' =0$ implies $w(s) = c+ds$; the 
eigenfunctions here are the restrictions of linear functions on $\RR^3$ to 
$S^2$, and so this gives the other family. It remains to show that these 
are the only temperate solutions.

It is clear that when $j\geq 2$, hence $\lambda \geq 6$, 
all solutions of $L_j'' w = 0$ grow 
exponentially in one direction or the other. On the other hand, consider the 
homogeneous system $L'_j$ when $j\geq 1$, which is
\begin{eqnarray*}
\frac{2}{3} u'' - \frac{\lambda}{2}u - \frac{\sqrt{\lambda}}{6} v' &=&0 \\
\frac{1}{2} v'' - (\frac{2}{3}\lambda-1)v + \frac{\sqrt{\lambda}}{6} u' &=&0,
\end{eqnarray*}
where $\lambda = j(j+1)$. This is a constant coefficient system, and 
solutions have the form 
\[
\left(\begin{array}{c} u \\ v \end{array}\right) = e^{\rho s}\,z_1, 
\qquad \mbox{or}\qquad se^{\rho s}z_1 + e^{\rho s}z_2, \qquad z_1,z_2 \in \C^2,
\]
where $\rho$ is a root of the indicial equation
\[
\det \left(\begin{array}{cc}\frac{2}{3}\rho^2 - \frac{\lambda}{2} &  
-\frac{\sqrt{\lambda}}{6}\rho \\ & \\
\frac{\sqrt{\lambda}}{6}\rho & \frac{1}{2}\rho^2 -(\frac{2}{3}\lambda-1)
\end{array}\right) = \frac{1}{3}(\rho^4 + 2(1-\lambda)\rho^2 +
(\lambda^2-\textstyle\frac{3}{2}\lambda)) = 0.
\]
Solutions are temperate only when $\mbox{Re}\,\rho = 0$, but it is
easy to check that this polynomial for $\rho$ has no root with
purely imaginary part, since $\lambda=j(j+1)$, $j \geq 1$. 
The proof is complete. 
\hfill $\Box$

\begin{remark}The space of conformal Killing fields on $S^3$ is 
10-dimensional, and is generated by the 6-dimensional family of rotations 
and the 4-dimensional family of dilations. Amongst all of these, only the 
vector fields vanishing at the two antipodal points (corresponding to the 
infinite ends on the cylinder) give rise to temperate conformal Killing 
fields on $\RR \times S^2$. Of course, these are just the 
generators for the 3-dimensional 
family of rotations of the orthogonal $S^2$ (i.e. corresponding to 
$\delta_{\theta}*\phi_i$, $i=1,2,3$, as above) and the dilation fixing 
these antipodal points ($\partial_s$). 
\end{remark}

\subsubsection{The lowest eigenvalue of $L$ on $[-T/2,T/2]\times S^2$}

The next application concerns the lowest eigenvalue of the operator 
$L$ with Dirichlet boundary conditions $X=0$ on the finite piece
of the cylinder $C_T = [-T/2,T/2] \times S^2$. 

\begin{proposition} Let $\lambda_0 = \lambda_0(T)$ denote the first Dirichlet 
eigenvalue for $L$ on $C_T= [-T/2,T/2] \times S^2$. Then 
\[
\lambda_0(T) \geq \frac{C}{T^2}
\] 
for some constant $C$ independent of $T$. 
\end{proposition}
{\bf Proof:} This estimate is clearly sharp for the scalar ordinary 
differential operators $L_0 = -\frac{2}{3}\del_s^2$ and $L''_1 = 
-\frac{1}{2}\del_s^2$. On the other hand, when $j\geq 2$, the lowest 
eigenvalue of $L''_j$ converges to $\frac{1}{2}j(j+1)-1$ as $T \to
\infty$, hence is bounded away from zero.

We finally show that the lowest eigenvalue for $L'_j$ is also bounded 
away from zero when $j\geq 1$. In fact,
\[
\bigg\langle L'_j\left( \begin{array}{c} u \\ v \end{array}\right),
\left( \begin{array}{c} u \\ v \end{array}\right)\bigg\rangle = 
\int_{-T/2}^{T/2} \frac{2}{3} (u')^2 + \frac{1}{2}(v')^2 + 
\frac{\sqrt{\lambda}}{6}(uv' -u'v) + \frac{\lambda}{2}u^2
+ (\frac{2}{3}\lambda - 1)v^2\, ds
\]
\[
= \int_{-T/2}^{T/2} \frac{2}{3}\left(u' - \frac{\sqrt{\lambda}}{8}v\right)^2
+ \frac{1}{2}\left(v' + \frac{\sqrt{\lambda}}{6}u\right)^2
+ \frac{35}{72}\lambda u^2 + (\frac{63}{96}\lambda-1)v^2\, ds.
\]
Since $\lambda \geq 2$, this is bounded below by $\textstyle\frac{1}{2}
\int_0^T(u^2+v^2)\, ds$, and the proposition is proved. \hfill $\Box$

\subsubsection{Growth estimates on $C_T$}

To conclude this analysis of $L$ on the cylinder, we show that if $X$ 
is an eigenfunction with `very small' eigenvalue, then its size in
the interior of $C_T$ is controlled by its size on the ends of $C_T$. 
The next result does this directly for the product metric. After that
we prove a more technical estimate for the inhomogeneous eigenvalue
problem which will be used later to handle the general case by perturbation. 

\begin{proposition} Suppose that $LX = \mu X$, where $L$ is computed
with respect to the product metric and $\mu < \frac18 \lambda_0(T)$. 
Suppose also that
\[
\int_{S^2} (|X(-T/2,\theta)|^2 + |X(T/2,\theta)|^2)\, d\theta \leq C_1.
\]
Then for any $a \in [-T/2+1,T/2-1]$,
\[
\int_{a-1}^{a+1}\int_{S^2} |X(s,\theta)|^2\, ds d\theta \leq C_2,
\]
where $C_2$ depends only on $C_1$, but not on $T$ or $a$, and where
$C_2\to 0$ as $C_1\to 0$.
\label{pr:supest}
\end{proposition}
{\bf Proof:} It suffices to check this estimate on each of the components 
in the decomposition of $L$. For the scalar operators $L_0$ and $L''_j$ 
this estimate may be verified using the explicit solutions of the
equation (though this can easily be done by other standard and less 
computational methods too). So as before, it suffices to focus on the 
system $L'_j$, $j \geq 1$. Write 
\[
L'_j = - A \del_s^2 + B \del_s + C
\]
where
\[
A = \left(\begin{array}{cc} 2/3 & 0 \\ 0 & 1/2 \end{array}\right),
\quad 
B = \left(\begin{array}{cc} 0 & \sqrt{\lambda}/6 \\ -\sqrt{\lambda}/6 & 0 
\end{array}\right), \quad
C = \left(\begin{array}{cc} \lambda/2 & 0 \\ 0 & ((2/3)\lambda - 1)
\end{array}\right),
\]
and, as usual, $\lambda = j(j+1)$. 

Now write $X = (u,v)^t$ and define $f(s) = \langle AX,X \rangle$. Then
\[
\frac{1}{2}\del_s^2 f(s) = \langle A X'',X\rangle + \langle A X',X'\rangle.
\]
Using the equation $L'_j X=\mu X$ and the skew-symmetry of $B$, this becomes
\[
\frac{1}{2}\del_s^2 f(s) = 
-\langle X',B X\rangle + \langle CX,X \rangle + \langle AX',X' \rangle
- \mu \langle X, X \rangle, 
\]
which can then be rewritten as
\[
\frac{1}{2}\del_s^2 f(s) = 
|A^{1/2}X' - \frac12 A^{-1/2}BX|^2 + \langle DX ,X \rangle,
\]
where
\[
D = \frac{1}{4}BA^{-1}B + C - \mu I = \left( \begin{array}{cc}
\frac{35}{72}\lambda - \mu & 0 \\ 0 & \frac{63}{96}\lambda - 1 - \mu 
\end{array}\right).
\]
But $D$ is positive definite when $\lambda \geq 2$ and $T$ is large 
enough that subtracting $\mu$ does not destroy positivity, and so we 
conclude that $f''(s) \geq 0$, and hence that the $L^2$ norm of $X$ 
at $s=\pm T/2$ controls the $L^2$ norm of $X$ over any strip $a-1 \leq s 
\leq a+1$.  \hfill $\Box$

\medskip

Our second result here is more specialized, and is the precise additional
ingredient required later to handle the nonproduct case. We let 
$\|\cdot \|_{0,\alpha,\sigma}$ denote the $\calC^{0,\alpha}$ norm over
the set $[\sigma,\sigma+1] \times S^2$. 

\begin{proposition} Let $L$ be computed relative to the product metric 
on $C_T$, and suppose $LX = \mu X + F$, where $X(\pm T/2) = 0$ and 
$\mu T^2< \nu_0$ for some sufficiently small constant $\nu_0$ independent
of $T$. Suppose also that for all $|s| \leq T/2$,
\[
\|F(s,\theta)\|_{0,\alpha,\sigma} \leq f(\sigma)\, e^{-T/2} \cosh \sigma,  
\]
where $|f(\sigma)| \leq 1$ for all $\sigma$.  Recalling that $F$ 
implicitly depends 
on $T$, we also assume that for any $\e > 0$ and $A>0$, there 
exists a $T_0$ such that when $T \geq T_0$ we have $|f(\sigma)| \leq \e$ 
when $T/2-A \leq |\sigma| \leq T/2$. Then, for any $\eta>0$ we have
$|X| \leq \eta$ on all of $C_T$, when $T$ is large enough.
\label{pr:supestgen}
\end{proposition}

In other words, if $X$ solves this inhomogeneous eigenvalue problem
on $C_T$, with vanishing Dirichlet data, if $\mu$ is much smaller than 
$1/T^2$, and if the forcing term $F$ is uniformly small near the ends of 
$C_T$ and bounded by $e^{-T/2}\cosh s$ on the middle portion of this 
cylinder, uniformly as $T \to \infty$, then $X$ is uniformly small on 
all of $C_T$.
\smallskip

\noindent {\bf Proof:} We separate variables and write the equation
as $L_j X_j = F_j$ for $j \geq 0$. For $j > 0$ this equation splits
further into $L_j' X_j' = F_j'$ and $L_j'' X_j'' = F_j''$, as
in the previous subsection. None of the primes here correspond
to derivatives.

We have shown that the operators $L_j'$ with $j \geq 1$ or $L_j''$ with 
$j \geq 2$ have spectrum bounded below by a positive constant
as $T \to \infty$. For any of the corresponding equations, if the
conclusion were false then we would be able to produce, by the
usual arguments, a nontrivial bounded function $X'$ or $X''$
such that $L_j' X' = 0$ or $L_j'' X'' = 0$, respectively, defined
on the complete cylinder $\RR \times S^2$. This contradicts the
strictly positive lower bound for the spectrum.

For the remaining components, it suffices to prove the assertion
for the scalar problem $\del_s^2 u + \alpha^2 u = F$, where $\alpha T$
is sufficiently small. 
For notational convenience we shall use the shifted variable
$t = s+T/2$ (so that $F(t) = f(t) e^{-T/2}\cosh (t-T/2)$, where
this function $f$ is a translate of the one above. We may write
the solution explicitly:
\[
u(t) = \frac{\sin \alpha(T-t)}{\sin \alpha T}
\int_0^t \frac{\sin \alpha \tau}{\alpha T}F(\tau)\, d\tau
+ \frac{\sin \alpha t}{\alpha T}
\int_t^T \frac{\sin \alpha(T-\tau)}{\sin \alpha T}F(\tau)\, d\tau.
\]
Because $\alpha T$ is so small, we may approximate each of the sines with 
their first order Taylor approximations; using the estimate for $F$ 
too we obtain
\[
|u(t)| \leq C \left( \left(1-\frac{t}{T}\right)\int_0^t \tau
\cosh (\tau - T/2)e^{-T/2} f(\tau)\, d\tau \right. \qquad \qquad \qquad
\]
\[
\qquad \qquad \qquad \left.
+ t \int_t^T \left(1 - \frac{\tau}{T}\right)\cosh(\tau - T/2)
e^{-T/2} f(\tau)\, d\tau\right).
\]
Finally, we must use the separate upper bounds for $f(\tau)$ in the 
various regions $0 \leq \tau \leq A$, $A \leq \tau \leq T-A$ and $T-A 
\leq \tau \leq T$. The main fact used in this estimation is that the 
function $h(t) = t(1-t/T)|\sinh(t-T/2)|e^{-T/2}$ is uniformly bounded 
independently of $T$, attains its maximum near $t=1$ and $t=T-1$,
and for any $\delta > 0$ there exists an $A>0$ such that
$h(t) \leq \delta$, uniformly in $T$, when $A \leq t \leq T-A$.
Together with the bounds for $f$, we can derive that $u$ is uniformly
small. Details are lengthy but straightforward, and are left to the reader.
\hfill $\Box$

\subsection{The lowest eigenvalue of $L$ on $\ST$} 

\begin{theorem} The lowest eigenvalue $\lambda_0 = \lambda_0(T)$
for $L$ on $\ST$ satisfies $\lambda_0 \geq CT^{-2}$ for some
constant $C > 0$ which is independent of $T$. 
\label{thm:egnval}
\end{theorem}

\noindent 
{\bf Proof of Theorem~\ref{thm:egnval} when the restriction of $\gamma_T$
to $C_T$ is the product metric:} 
If this result were false, it would be possible to find a 
sequence of values $T_j \to \infty$ such that for the operators $L_{T_j}$ 
associated to the metrics $\gamma_{T_j}$, we would have $\lambda_0(T_j)
T_j^2 \to 0$. We now show this leads to a contradiction. 

Denote the sequence of metrics and operators by $\gamma_j$ and $L_j$,
respectively. Let $X_j$ be any one of the eigenfunctions for $L_j$ 
corresponding to the eigenspace with eigenvalue $\lambda_0(T_j)$. 
By rescaling, assume that $\sup_{\Sigma_{T_j}} |X_j| = 1$. 

The coefficients of $L_j$ converge uniformly on any compact subset of 
$\Sigma^*$ to the coefficients of the operator $L_c$ corresponding
to the metric $\gamma_c$. Since $|X_j|$
is uniformly bounded, we may extract a subsequence $X_{j'}$ which 
converges uniformly on compact sets in $\Sigma^*$ to a limit $X$.
This vector field is a solution of $L_c X = 0$ and satisfies
$|X| \leq 1$. 

Our first claim is that $X$ is nontrivial. To see this, relabel the 
subsequence as $X_j$ again. Suppose that the supremum of $|X_j|$ 
is attained at some point $q_j \in \Sigma_{T_j}$. If $q_j$ lies in 
$\Sigma_{T_j} \setminus C_{T_j}$ for infinitely many $j$, then 
again possibly passing to a subsequence $q_j \to q \in \Sigma^*$. 
Since $|X_j(q_j)| \equiv 1$, we get $|X(q)| = 1$, and so $X \not \equiv 0$. 
On the other hand, suppose that $q_j \in C_{T_j}$ for all but finitely many 
$j$. Note that by Schauder theory, $|X_j|$ remains bounded away from zero 
on a strip $[a_j-\eta,a_j+\eta]\times S^2 \subset C_{T_j}$ of width
not depending on $j$, about $q_j$. 
Furthermore, $\lambda_0(T_j) < \frac18 T_j^2$ for $j$ large enough,
and hence Proposition~\ref{pr:supest} (and elliptic theory) implies that 
$|X_j(q_j')| \geq \alpha > 0$ for some point $q_j' \in \del C_{T_j}$
and some $\alpha$ not depending on $j$. Therefore in this case
too we may conclude that the limit $X \not \equiv 0$ on $\Sigma^*$.

By the analysis of \S 3.1, if $X$ is any solution of $L_c X = 0$ on 
$\Sigma^*$ (with respect to the metric $\gamma_c$) which is bounded along 
the cylindrical end, then it can be written as a sum $X = X_\infty + Z$, 
where $X_\infty$ extends to a bounded solution of $L X_\infty=0$ on the 
entire cylinder $\RR \times S^2$ (with product metric) and 
$Z$ decays exponentially as $t \to \infty$. 

Our second claim is that, assuming the nondegeneracy of $\Sigma$ so that 
there are no conformal Killing fields vanishing at both points $p_j$, 
then any bounded vector field $X$ on $\Sigma^*$ which satisfies
$L_c X = 0$ necessarily vanishes identically. To see this, observe
from our earlier result about elements in the nullspace of $L_c$ which
are bounded on the cylinder, all such solutions are also annihilated by 
$\D$. (This is not true if we also admit the linearly growing
solutions.) Thus the term $X_\infty$ in the decomposition of $X$ above 
is annihilated by $\D$, and hence $\D X = \D Z$ decays exponentially
along the cylindrical ends. Therefore, no boundary contribution
appears in the integration by parts
\[
0 = \langle L_c X,X\rangle = \langle \D^* \D X,X \rangle = ||\D X||^2,
\]
so that $\D X \equiv 0$. Thus $X$ is a conformal Killing field on 
$\Sigma^*$ which is bounded with respect to the metric $\gc$;
changing back to the old polar variables $r_j = e^{-t_j}$ and $\theta$
on each ball $B_j$ shows that $X$ is bounded with respect to the original 
metric $\gamma$ on $\Sigma$, and in fact vanishes at both points $p_j$. 
It is therefore in the (distributional) nullspace of $L$ on $\Sigma$, 
and hence by elliptic regularity extends smoothly to a solution on all 
of $\Sigma$. It would thus be a nontrivial conformal Killing field on the 
manifold $\Sigma$, vanishing at both $p_1$ and $p_2$, and we have assumed 
that such a field does not exist. This completes the proof.
\hfill $\Box$

\medskip

\noindent {\bf Proof of Theorem~\ref{thm:egnval} in general:}
We may follow the outline of the preceding proof at almost
every point. Thus we suppose the result false and choose
a sequence of eigenfunctions $X_j$ with corresponding very
small eigenvalues $\mu_j$, for some sequence $T_j \to \infty$.
Supposing that $|X_j| \leq 1$, we again attempt to pass to
a limit and obtain a contradiction. If $|X_j|$ remains
bounded away from zero on some fixed compact set near
the ends, then we argue as before. The difficulty, however,
is when $|X_j|$ remains bounded away from zero in some
portion of the cylinder $C_{T_j}$ which recedes from either
end.  To handle this, we would like to use the same sort of
argument as in the proof for the product case. Namely, we would
like to assert that if $|X_j|$ tends to zero near the ends
of $C_{T_j}$, then it must tend to zero uniformly along
all of this cylindrical segment. We shall now prove this
using Proposition~\ref{pr:supestgen}.

The metric $\gamma_T$ differs from the product metric on $C_T$ by a term 
of size $e^{-T/2}\cosh s$, and therefore the operator $L$ computed relative 
to $\gamma_T$ differs from the corresponding operator $L_0$, computed 
relative to the product metric, by an error term $E_T$ with coefficients
of this same size. We may rewrite the equation for $X_j$ as
\[
L_0 X_j = \mu_j X_j + F_j,
\]
where $F_j = -E_{T_j}X_j$. Now suppose that we are in the difficult situation: 
that $|X_j|$ is uniformly small on any fixed neighborhood of the ends of 
the cylinder. Then it is clear that $F_j$ satisfies precisely the estimates 
in the statement of Proposition~\ref{pr:supestgen}. The solution
$X_j$ may be decomposed into a sum $U_j + V_j$, where $U_j$ satisfies
the homogeneous problem $L_0 U_j = \mu_j U_j$ and $U_j(\pm T/2) =
X_j(\pm T/2)$, while $V_j$ satisfies the inhomogeneous problem with
vanishing boundary conditions. Clearly both of these summands are
uniformly small near the ends of $C_{T_j}$. The uniform smallness
of $|U_j|$ on the entire cylinder now follows from Proposition~\ref{pr:supest},
while the uniform smallness of $|V_j|$ follows from 
Proposition~\ref{pr:supestgen}. Therefore it is impossible that $X_j$ 
is large in the middle of $C_{T_j}$ but small at the ends. 
This completes the proof. \hfill $\Box$

\subsection{Estimates on H\"older spaces}

We have shown that the lowest eigenvalue of the operator $L$ 
on $\ST$ is bounded below by $CT^{-2}$, uniformly as $T \to \infty$.
Hence the norm of $G = L^{-1}$, as a bounded operator on $L^2(\Sigma_T)$,
blows up no faster than $C'T^2$. However, it is more convenient
for us to use H\"older spaces because of the nonlinear nature
of the problem, and we show now that this $L^2$ estimate may
be converted to a slightly weaker estimate on appropriately
defined H\"older spaces. 

\begin{definition}
Consider the cover of $\Sigma_T$ given by the sets $\Sigma^*_{R/2}$
and $C_T$, and for any vector field $X$ on $\Sigma_T$, let
$X'$ and $X''$ denote the restrictions of $X$ to these two
subsets, respectively. Also, for $-T/2+1 \leq a \leq T/2-1$,
define $N_a = [a-1,a+1] \times S^2 \subset C_T$. Then for any $k 
\in {\mathbb N}$ and $\alpha \in (0,1)$, define
\[
||X||_{k,\alpha} = ||X'||_{k,\alpha} + 
\sup_{-T/2+1 \leq a \leq T/2-1} ||X''||_{k,\alpha, N_a}.
\]
The local H\"older norms here are computed with respect to
the metric $\gamma_T$. 
\end{definition}

\begin{corollary} Let $W \in {\mathcal C}^{k,\alpha}(\Sigma_T)$
and suppose that $X$ is the unique solution to $L_{}X = W$, as 
provided by Theorem~\ref{thm:egnval}. Then for some
constant $C$ independent of $W$ and $T$, 
\[
||X||_{k+2,\alpha} \leq C T^3 ||W||_{k,\alpha}.
\]
\label{co:fhest}
\end{corollary}
{\bf Proof:} First, by Theorem \ref{thm:egnval}, $||X||_{L^2} \leq C 
T^2 ||W||_{L^2}$. Next, since the volume of $\Sigma_T$ is 
bounded by $CT$, we have $||W||_{L^2} \leq T ||W||_{0,\alpha}$. 
In addition, local Schauder estimates for $L_{c}$ give $||X||_{2,\alpha} 
\leq C(||W||_{0,\alpha} + ||X||_{L^2})$. Putting these together yields
\[
||X||_{2,\alpha} \leq CT^3 ||W||_{0,\alpha},
\]
which is the desired estimate when $k = 0$. The estimate when
$k > 0$ requires another application of the local Schauder estimates. 
\hfill $\Box$ 

\subsection{Correcting $\mu_T$ to be transverse-traceless}
Recall now that $\mu_T$ is constructed by patching together
$\mu_c$ using cutoff functions in the center $Q$ of $C_T$,
and the metric $\gamma_T$ is constructed from $\gamma_c$ the
same way. The definition (\ref{muic-size}) implies that on $Q$
\begin{eqnarray}
|\mu_T|_{\gamma_T} \leq C \psi_T^{6}|\mu|_{\gamma}
\label{eq:sizemut}
\end{eqnarray}
with equality (for $C=1$) on $\ST\setminus Q$.
Since $\mu_c$ is divergence-free with respect to $\gc$, the
`error term' $\dive_{\gamma_T}\mu_T$ arises from two sources: the cutoff 
functions used to define $\mu_T$ and those used to define $\gamma_T$. 
This error term is clearly supported in the center $Q$ of $C_T$,
and it is not hard to see that
\[
||\dive_{\gamma_T}\mu_T||_{k-1,\alpha} \leq C' ||\mu_T||_{k,\alpha} \leq 
C'' ||\psi_T^6||_{k,\alpha},
\]
the last inequality following from (\ref{eq:sizemut}) since 
$|\mu|_\gamma \leq C$. Recalling that $\psi_T \sim e^{-t_j/2}$ and 
each $t_j \sim T/2$ on $Q$, and letting $W$ be the vector field dual to 
$\dive_{\gamma_T}\mu_T$, then we conclude that
\begin{equation}
||W||_{k+1,\alpha}\leq C\, e^{-3T/2}.
\label{eq:3.6}
\end{equation}

\begin{proposition}
\label{tt-prop}
There exists a tensor $\sigma_T$ on $\Sigma_T$ with 
\begin{equation}
||\sigma_T||_{k,\alpha}\leq C\, T^3 e^{-3T/2}
\label{eq:sizemutt}
\end{equation}
and such that $\tilde{\mu}_T = \mu_T - \sigma_T$ is 
transverse-traceless with respect to $\gamma_T$. 
\end{proposition}
{\bf Proof:} As indicated in the beginning of this section, 
we first solve the equation $L X = W$ on $\Sigma_T$, with 
$W = \dive_{\gamma_T} \mu_T$, and then set
$\sigma_T = \D X$. The operators $\D$ and $L$ here
are both computed with respect to $\gamma_T$. The estimates
for $\sigma_T$ and $\tilde{\mu}_T$ follow from (\ref{eq:3.6}). 
\hfill $\Box$

\section{Estimates on the approximate solution}
\label{se:4}

Our overall goal in this paper is to modify the approximate 
initial data set $(\gamma_T,\mu_T,\psi_T)$, defined on the
manifold $\Sigma_T$, to a genuine initial data set
$(\gamma_T,\tilde{\mu}_T,\tilde{\psi}_T)$, so that
$\tilde{\mu}_T$ is transverse-traceless with respect to $\gamma_T$
and $\tilde{\psi}_T$ satisfies the Lichnerowicz equation
relative to $\gamma_T$ and $\tilde{\mu}_T$. Thus far we have
accomplished the first of these tasks, and so what remains is to 
show how to modify the conformal factor $\psi_T$ to $\tilde{\psi}_T = 
\psi_T + \eta_T$ so that $\tilde{\psi_T}$ satisfies the
Lichnerowicz equation. There are three separate steps
involved in this. In the first we must calculate the `error term'
$E_T$, which measures the deviation of $\psi_T$ from solving
the Lichnerowicz equation. After that we must analyze the
mapping properties of the linearization of this equation,
calculated at $\psi_T$, uniformly as $T \to \infty$. Finally,
these are put together to carry out the contraction mapping
argument to produce the exact solution $\tilde{\psi}_T$. 
These steps are taken up in this and the next two sections,
successively.

To begin then, denote by $\calN_T$ the Lichnerowicz operator
with respect to $(\gamma_T,\tilde{\mu}_T)$:
\begin{equation}
\calN_T(\psi) = \Delta_T \psi - \frac18 R_T \psi +
\frac18 |\tilde{\mu}_T|^2 \psi^{-7} - \frac{1}{12}\tau^2 \psi^5.
\label{eq:licht}
\end{equation}
The Laplacian and scalar curvature and norm squared of $\tilde{\mu}_T$
are all computed with respect to $\gamma_T$ here. We define the
error term
\begin{equation}
E_T = \calN_T(\psi_T),
\label{eq:et}
\end{equation}
which measures the deviation of $\psi_T$ from being an exact
solution.

\begin{proposition}
For every $k$ and $\alpha$ there exists a constant $C$ which does
not depend on $T$ such that $||E_T||_{k,\alpha} \leq C e^{-T/2}$.
\label{pr:errest}
\end{proposition}

The proof of this proposition involves a careful accounting
of the various terms in $\calN_T(\psi_T)$.

We begin by noting that away from the neck region $C_T$,
the estimate is nearly trivial. In fact, on $\Sigma_R^*$,
$\gamma_T = \gamma_c$, $\psi_T = \psi_c$ and $\tilde{\mu}_T
= \mu_c + {\mathcal O}(T^3 e^{-3T/2})$. Inserting these
into the Lichnerowicz equation, and using that $(\gamma_c,
\mu_c,\psi_c)$ is an exact solution, we have that
\[
|E_T| \leq C T^3 e^{-3T/2} \leq C e^{-T/2},
\]
along with all its derivatives, as claimed. 

We divide the tube $C_T$ up into three regions: the center
$Q = [-1,1]\times S^2$, and the regions to the left
and right of this, $[-T/2,-1]\times S^2 \equiv C_T^{(1)}$
and $[1,T/2] \times S^2 \equiv C_T^{(2)}$.
We also recall the notation used in the construction of
the approximate solution that $\gamma_j$, $\mu_j$ and
$\psi_j$ denote the restrictions of $\gamma_c$, $\mu_c$
and $\psi_c$ to the balls $B_j$, and hence, by a small
abuse of notation, to the tube $C_T$ as well. Thus
$(\gamma_1,\mu_1,\psi_1)$ represents data `coming in
from the left', while $(\gamma_2,\mu_2,\psi_2)$
represents data `coming in from the right'. Recall also
that $\gamma_T$ and $\mu_T$ agree with $\gamma_j$ and
$\mu_j$ on $C_T^{(j)}$, whereas $\psi_T$ is a sum
of two terms, one much larger than the other, in each
of these side regions.

We begin by listing some trivial estimates which hold in all of $C_T$, 
and which follow from the construction of the approximate solution
and from (\ref{eq:sizemutt}):
\begin{itemize}
\item[a)] $\gamma_T = ds^2 + h + {\mathcal O}(e^{-T/2}\cosh s)$,
\item[b)] $\psi_T \sim 2e^{-T/4}\cosh (s/2)$,
\item[c)] $|\mu_T| \sim \psi_T^6 \leq C e^{-3T/2}(\cosh (s/2))^6$,
\item[d)] $|\tilde{\mu}_T - \mu_T| \leq C T^3 e^{-3T/2}$.
\end{itemize}
We shall also repeatedly use the fact that
\begin{equation}
\Delta_{\gamma_j} \psi_j - \frac18 R(\gamma_j)\psi_j
+ \frac18 |\mu_j|^2 \psi_j^{-7} - \frac1{12}\tau^2 \psi_j^5 \equiv 0.
\label{lich_i}
\end{equation}

We begin by estimating $E_T$ in $Q$.  The main observation
concerning this estimate is that we have a decomposition
\[
\pt=\psi_1+\psi_2
\]
valid in $Q$. In addition, it follows from property (a) above
and from the definition of $\gt$ that 
\[
\Delta_T - \frac18 R_T = \Delta_{\gamma_j} - \frac18 R_{\gamma_j} + 
{\mathcal O}(e^{-T/2})
\]
for $j=1$ or $2$.
On $Q$, we may therefore expand $(\Delta_T - \frac18
R_T)\psi_T$ as a sum of two terms. Since on $Q$,
$\psi_j={\mathcal O}(e^{-T/4})$, we see from (\ref{lich_i}) that
we have 
\[
(\Delta_T - \frac18 R_T)\psi_T = \sum_{j=1}^{2}
(-\frac18 |\mu_j|^2 \psi_j^{-7} + \frac1{12}\tau^2 \psi_j^5)
+ {\mathcal O}(e^{-3T/4}). 
\]
From this, using the estimate that 
$|\mu_j|^2={\mathcal O}(e^{-3T})$ in $Q$, we easily conclude that
\[
(\Delta_T - \frac18 R_T)\psi_T = {\mathcal O}(e^{-3T/4}). 
\]
As for the other terms in the Lichnerowicz equation, using 
properties (b), (c) and (d) above we easily get suitable
estimates for them in this region as well. In particular, we have 
$|\tilde{\mu}_T|^2 \psi_T^{-7} = {\mathcal O}(T^6 e^{-5T/4})$ and
$\tau^2\pt^5 = {\mathcal O}(e^{-5T/4})$. Hence in $Q$, we obtain 
$|E_T|= {\mathcal O}(e^{-3T/4})$, which is even better than the estimate
we have stated.

The estimates in $C_T^{(1)}$ and $C_T^{(2)}$ are nearly identical
to one another, so it suffices to concentrate on just one of these,
let us say in $C_T^{(2)}$ to be concrete.
In this region, $\psi_T = \psi_2 + \chi \psi_1$, where
$\chi$ is a function cutting $\psi_1$ off to zero at the
far end of $C_T$, where $s \approx T/2$. In this region
\[
\psi_1 = e^{-T/4}e^{-s/2} + {\mathcal O}(e^{-T/4}e^{-3s/2}), 
\qquad \psi_2 \sim e^{-T/4}e^{s/2},
\]
and in addition 
\[
\Delta_T - \frac18 R_T = \Delta_0 - \frac18 R_0 + 
{\mathcal O}(e^{-T/2}\cosh s).
\]
Therefore 
\[
(\Delta_T - \frac18 R_T)(\chi \psi_1) = e^{-3T/4}\cosh s e^{-s/2}
+ {\mathcal O}(e^{-T/2}) = {\mathcal O}(e^{-T/2}).
\]
Next, $\psi_T = \psi_2(1 + \chi \psi_1/\psi_2)$ and hence
\[
\psi_T^{-7} = \psi_2^{-7} + {\mathcal O}(e^{7T/4}e^{-9s/2}),
\]
and
\[
\psi_T^5 = \psi_2^5 + {\mathcal O}(e^{-5T/4}e^{3s/2}).
\]
Furthermore, we have here
\[
|\tilde{\mu}_T|^2 = |\mu_2|^2 + {\mathcal O}(T^3 e^{-3T/2}) 
\]
Putting these estimates all together yields $|E_T| 
\leq Ce^{-T/2}$ on $C_T^{(2)}$. With a similar estimate on  $C_T^{(1)}$
we conclude that $|E_T|\leq Ce^{-T/2}$ on all of $C_T$.
The estimates on the derivatives of $E_T$ follow similarly.

\section{The linearization of the Lichnerowicz operator} 
\label{se:5}

Our main goal now is to find a correction term $\eta_T$ for the 
approximate solution $\psi_T$ so that the sum $\psi_T + \eta_T$ solves 
the Lichnerowicz equation ${\mathcal N}_T(\psi_T +\eta_T) = 0$
on $\Sigma_T$. This will be done in the next section using
a contraction mapping argument. In preparation for this we establish 
in this section good estimates for the mapping properties of the 
linearization $\calL_T$ of this operator $\calN_T$ about the 
approximate solution $\psi_T$; as usual, keeping track of
behavior as $T \to \infty$. 

We first record that the linearization $\calL$ of the Lichnerowicz 
operator on $\Sigma$ with respect to the initial data set $(\gamma,\mu)$ 
is given by 
\[
\calL = \Delta_{\gamma} -\frac18\left( R(\gamma)
+7\,|\mu|^{2}_{\gamma} + \frac{10}{3} \,\tau^2
\right),
\]
but using the fact that the pair $(\gamma,\mu)$ satisfies the 
constraint equations, we can rewrite this as 
\begin{equation}
\calL = \Delta_{\gamma} -\left( |\mu|^{2}_{\gamma} + 
\frac{1}{3} \,\tau^2\right).
\label{eq:4.3}
\end{equation}
By standard elliptic theory (for $\Sigma$  compact),  
\[
{\cal L} : {\mathcal C}^{k+2,\alpha}(\Si) \rightarrow  
{\mathcal C}^{k,\alpha}(\Si) 
\]
is a Fredholm mapping of index zero for every $k \in {\mathbb N}$ and 
$\alpha\in (0,1)$. Furthermore, since we are assuming that $\Pi \not\equiv 
0$, which is equivalent to $\tau \neq 0$ or $\mu\not\equiv 0$, 
the term of order zero here is nonpositive and not identically zero
and so the maximum principle implies that there are no nontrivial
solutions of $\calL \phi = 0$. Consequently the cokernel is also 
trivial, and we see that (\ref{eq:4.3}) is always an isomorphism.

Our first goal in this section is to establish that the same property is 
true for the linearization $\calL_T$ on $\Sigma_T$, which is given by
\begin{equation}
{\cal L}_{T} = \Delta_{\gamma_T} -\frac{1}{8}\left( R(\gamma_T)
+7\,|\tilde{\mu}_T|^{2}\,\pt^{-8} + \frac{10}{3} \,\tau^2\,\pt^4
\right).
\label{eq:4.1}
\end{equation}
This operator is elliptic and Fredholm of index zero acting on the 
H\"older spaces introduced in \S 3.3; unfortunately, the term of 
order zero here is no longer necessarily nonpositive since it can no 
longer be rewritten using the constraint equations, and we do
not have control on the sign of $R(\gamma_T)$ near the
boundary of $\Sigma_R^*$. Therefore, the triviality of the nullspace 
of $\calL_T$ when $T$ is large enough requires further argument. 
We must also show that the norm of the inverse 
operator is uniformly bounded as $T \to \infty$.  This last issue 
is complicated by the following considerations. We shall be using 
this solution operator in an iterative scheme to produce the correction 
term $\eta_T$ for the approximate solution $\psi_T$. Because of the term 
$\psi^{-7}$ in the Lichnerowicz equation, it 
is quite important that $\eta_T$ be substantially smaller than
$\psi_T$. Furthermore, we also desire that $\eta_T$ vanish 
exponentially as $T \to \infty$ on the `body' $\Sigma_R^*$, away
from the scene of the surgery. To ensure that both of these requirements
are met in the iteration scheme, we establish precise estimates for the 
solution operator for $\calL_T$. This necessitates using slight alterations 
of the H\"older spaces which incorporate a weight factor along the neck. 
Thus we first define these weighted H\"older spaces and then take up
the matters of the existence and uniformity of the inverse of $\calL_T$
on these spaces.

\begin{definition}
Let $w_T$ be an everywhere positive smooth function on $\Sigma_T$ 
which equals $e^{-T/4}\cosh(s/2)$ on $C_T$ and which is uniformly bounded 
away from zero on $\Sigma_R^*$. (We may as well also assume that $w_T 
\equiv 1$ on $\Sigma_{2R}^*$.) Now, for any $\delta\in \R$, and any $\phi
\in {\mathcal C}^{k,\alpha}(\Sigma_T)$, set 
\[
||\phi||_{k,\alpha,\delta} = ||w_T^{-\delta}\phi||_{k,\alpha},
\]
and let ${\mathcal C}^{k,\alpha}_\delta(\Sigma_T)$ denote the
corresponding normed space. 
\label{ct-weight}
\end{definition}

\begin{proposition}
Fix any $\delta\in\R$. For $T$ sufficiently large, the mapping 
\[
\calL_T : {\mathcal C}^{k+2,\alpha}_{\delta}(\Sigma_T) 
\longrightarrow  
{\mathcal C}^{k,\alpha}_{\delta}(\Sigma_T) 
\]
is an isomorphism.
\label{pr:2}
\end{proposition}
{\bf Proof:} The action of $\calL_T$ on ${\mathcal C}_\delta^{k+2,\alpha}
(\Sigma_T)$ is equivalent to the action of the conjugated operator 
$w_T^{-\delta}\calL_T w_T^{\delta}$ on the unweighted space 
${\mathcal C}^{k+2,\alpha}$. As $\delta$ varies, these mappings are all 
Fredholm of index zero. Furthermore, any element of the nullspace of $\calL_T$
is in any one of the spaces ${\mathcal C}^{k+2,\alpha}_\delta$.
Therefore, to prove that any one of these maps is an isomorphism, it 
suffices to show that the nullspace of $\calL_T$ is trivial when $\delta = 0$. 
In addition, by elliptic regularity, it even suffices to show that
there is no element $\eta \in {\mathcal C}^0(\Sigma_T)$ such that
$\calL_T \eta = 0$. 

Assume this is not the case, so that there exists a sequence $T_j
\to\infty$ and functions $\eta_j\in C^{0}(\Si_{T_j})$ satisfying
$\calL_{T_j}(\eta_j) =0$. Write $\calL_{T_j} = \calL_j$ and
normalize $\eta_j$ so that $\sup |\eta_j| = 1$. As in the proof
of Theorem~\ref{thm:egnval}, we shall show first that we may
extract a nontrivial limiting function $\eta$, and then show that
the existence of this function leads to a contradiction. 

Suppose first that the supremum of $|\eta_j|$ on $\Sigma_{R}^*$
remains bounded away from zero, at least for infinitely many $j$.
Then using local elliptic theory to control higher order
derivatives, we may extract a subsequence $\eta_{j'}$ which
converges to some limiting function $\eta \not\equiv 0$ on $\Sigma^*$.
Since the coefficients of $\calL_j$ converge on compact
sets to those of $\calL_c$, the linearization of the Lichnerowicz
operator with respect to the initial data set $(\gamma_c,\mu_c)$
at $\psi = \psi_c$, we see that 
\begin{equation}
\calL_c \eta = 
\Delta_{\gamma_c} \eta - \frac18 \left(R(\gamma_c)
+ 7 |\mu_c|^2\psi_c^{-8} + \frac{10}{3}\tau^2 \psi_c^4\right) \eta = 0.
\label{eq:4.655}
\end{equation}
The first two terms, $\Delta_{\gamma_c} - \frac18 R(\gamma_c)$,
constitute the conformal Laplacian for $\gamma_c$. Since 
$\gamma_c = \psi_c^{-4}\gamma$, the familiar conformal covariance 
property of this operator leads to 
\begin{equation}
\Delta_{\gamma_c}\eta -\frac{1}{8}\,R(\gamma_c)\,\eta= \psi_c^5
\left(\Delta_{\gamma}(\psi_c^{-1}\eta) -\frac{1}{8}\,R(\gamma)\,
\psi_c^{-1}\eta\right).
\label{eq:4.7}
\end{equation} 
Using this in (\ref{eq:4.655}), as well as the fact that 
$|\mu_c|_{\gamma_c}^2 = \psi_c^{12} |\mu|_{\gamma}^2$, and dividing
through by $\psi_c^5$, we conclude that $\calL u = 0$ on $\Sigma^*$,
where $u = \eta/\psi_c$. (Note that $\calL$ in this equation is the 
operator which extends smoothly across the $p_j$.) Next, $\eta$ 
is bounded on $\Sigma^*$ and $\psi_c \sim r_j^{1/2}$ near $p_j$, 
and so, in $B_j$, $|u| \leq C r_j^{-1/2}$. This singularity is weaker 
than that of the Green's function in three dimensions, and so $u$ extends
to a weak solution of $\calL u = 0$ on all of $\Sigma$. But we have
already remarked earlier that the operator $\calL$ on $\Sigma$ has 
trivial nullspace, which means that $u \equiv 0$. This is a contradiction.

It remains to address the case in which the supremum of $|\eta_j|$
on $\Sigma^*_{R}$ converges to zero. In fact, if $\sup |\eta_j|$ is
attained at some point $q_j \in \Sigma_{T_j}$, then we may
apply the same argument as before if $q_j$ remains within 
a bounded distance of the end of the cylindrical piece $C_{T_j}$. 
If this distance tends to infinity, then choose a new linear coordinate
$s'$ which is centered at the point $q_j$ (so that $s' = 0$
at that point) and is a translation of the coordinate  $s$. 
Using local elliptic theory again we may extract
a subsequence which converges on any compact set, along with all
its derivatives, to some nontrivial function $\eta$ defined
on the complete cylinder $\RR \times S^2$. Furthermore, the
metric $\gamma_{T_j}$ converges in this limit to the product
metric $\gamma_0 = (ds')^2 + h$, $R(\gamma_{T_j}) \to 2$, and the terms 
$|\mu_{T_j}|^2\psi_{T_j}^{-8}$ and $\psi_{T_j}^4$ both converge to zero.
Hence $\calL_j$ converges locally on compact sets to $\Delta_{\gamma_0}
- 1/4$. Thus we have obtained in this limit a nontrivial bounded
solution of the equation
\begin{equation}
\calL_{\gamma_0}\eta=\Delta_{\gamma_0}\eta-\frac{1}{4}\eta=0
\end{equation}
on the entire cylinder $\RR \times S^2$, and such a solution
obviously cannot exist. Hence this case too leads to a contradiction.

This completes the proof of Proposition~\ref{pr:2}.
\hfill$\Box$

\medskip

For $T$ sufficiently large, let $\calG_T$ denote the
inverse of $\calL_T$ acting on ${\mathcal C}^{k,\alpha}_\delta$. 
In other words, we have 
\begin{equation}
\calG_T: {\mathcal C}^{k,\alpha}_{\delta}(\ST)\longrightarrow 
{\mathcal C}^{k+2,\alpha}_{\delta}(\ST),
\label{eq:mapgt}
\end{equation}
with $\calL_T \calG_T = \calG_T \calL_T = I$. 
Although of course $\calG_T$ depends on the chosen weight $\delta$, we
suppress this dependence in the notation.

\begin{proposition}
If $0 < \delta < 1$, then the norm of the operator
${\calG}_T$ in (\ref{eq:mapgt}) is uniformly bounded as $T \to \infty$.
\label{pr:3}
\end{proposition}

{\bf Proof:} The strategy is the same as before. If the result
were false, then there would exist a sequence $T_j\to\infty$ and 
functions $f_j\in {\mathcal C}^{k,\alpha}_{\delta}(\Si_{T_j})$ such that,
writing $\calG_j = \calG_{T_j}$,
\[
||f_j||_{k,\alpha,\delta} \to 0 \qquad \mbox{and}
\qquad ||{\calG}_j(f_j)||_{k+2,\alpha,\delta}=1
\]
as $j \to \infty$. 

Setting $v_j = \calG_j (f_j)$, we argue exactly as in the preceding 
proof. Either some subsequence of the $v_j$ converges to 
a bounded, nontrivial function $v$ on $\Sigma^*$ which
satisfies $\calL_c v = 0$, which leads to a contradiction, or else 
some subsequence converges on the cylinder. This second case
is precisely why we have introduced the weight function 
$w_{T_j}^{\delta}$. 

Write $w_{T_j} = w_j$ and $C_{T_j} = C_j$. Since $||v_j||_{k+2,\alpha,
\delta} \equiv 1$, we have 
\[
1 \geq \sup_j \sup_{C_j} w_j^{-\delta}|v_j| \geq c > 0.
\]
Suppose that $\sup_{C_j}w_j^{-\delta}|v_j|$ is attained at 
some point $q_j \in C_{j}$. We may also assume that $|v_j| \to 0$ on 
any fixed compact set of $\Sigma^*$ away from $C_j$. Now, by assumption
$|v_j| \leq w_j^{\delta}$, and if we renormalize by setting 
$\tilde{v}_j = e^{T_j/4}v_j$, then this becomes
\[
|\tilde{v}_j(s,\theta)| \leq (\cosh (s/2))^{\delta}
\]
on $C_{T_j}$, with equality attained at the point $q_j$. Writing $q_j 
= (s_j,\theta_j)$, then we renormalize yet again, setting 
$\hat{v}_j(s',\theta) = \tilde{v}_j(s'+s_j,\theta)/(\cosh (s_j/2))^\delta$.
Hence
\[
|\hat{v}_j(s',\theta)| \leq \left(\frac{\cosh \frac 12(s'+s_j)}{
\cosh \frac12 s_j} \right)^\delta,
\]
with equality at the point $(s',\theta) = (0,\theta_j)$. In
particular, $\hat{v}_j(0,\theta_j) = 1$. Now 
we can pass to a limit. If $s_j \to \pm \infty$, then the
right hand side here converges to $e^{\delta s'/2}$ or $e^{-\delta s'/2}$,
while if $s_j$ remains bounded, then the right side converges to
a function which is bounded by $C (\cosh (s'/2))^\delta$ for
some $C>0$, while the left side converges to a nontrivial function 
$\hat{v}$ which 
satisfies $\calL_{\gamma_0}\hat{v} = 0$. But any solution of
this equation blows up at least like $e^{|s'|/2}$ either
as $s' \to \infty$ or as $s' \to -\infty$. Since $0 < \delta< 1$,
this rate of growth is incompatible with the previous inequalities, 
and so we reach a contradiction once again. \hfill$\Box$

\section{Proof of the main Theorem}
\label{se:6}

We wish to solve 
\begin{equation}
{\cal N}_{T}(\pt +\eta_T)=0.
\label{eq:5.1}
\end{equation}
Expanding ${\cal N}_{T}(\pt +\eta)$ in a Taylor series about $\eta=0$ we have
$$
{\cal N}_{T}(\pt +\eta)= E_T +{\cal
L}_{T}(\eta) + {\cal Q}_{T}(\eta),
$$
where $E_T$ and ${\cal L}_{T}$ are given by 
(\ref{eq:et}) and (\ref{eq:4.1}) respectively, and 
\begin{eqnarray*}
{\cal Q}_{T}(\eta) &=& {\cal N}_{T}(\pt +\eta)-E_T 
-{\cal L}_{T}(\eta)\\
&=& \frac{1}{8} |\tmut|_{\gt}^{2}\left( (\pt +\eta)^{-7} - \pt^{-7}
+ 7 \pt^{-8}\eta\right)\\ 
&&- \frac{1}{12}\tau^2\left( (\pt +\eta)^{5} 
- \pt^{5}
-5\pt^4 \eta\right)\\
&=&  \frac{1}{8} |\tmut|_{\gt}^{2}\pt^{-7}\left( (1
+\frac{\eta}{\pt})^{-7} - 1
+ 7 \frac{\eta}{\pt}\right)\\
&&- \frac{1}{12}\tau^2\left( 
10\pt^3\eta^2 + 10\pt^2\eta^3 + 5\pt\eta^{4} +\eta^{5}\right). 
\end{eqnarray*} 
is the ``quadratically vanishing'' nonlinearity.
If we require that $\eta\in {\mathcal C}^{k+2,\alpha}_{\delta}(\ST)$ 
satisfy $|\eta|< c\pt$ 
for some constant $c<1$, then from the expansion above 
we easily obtain an estimate of the form

\begin{equation}
\|{\cal Q}_{T}(\eta)\|_{k+2,\alpha,\delta}\leq C
\|\eta\|_{k+2,\alpha,\delta}^2
\label{quad-est}
\end{equation}
for some constant $C$ independent of $T$.
Equation (\ref{eq:5.1}) is equivalent to 
$$
\eta_T = -{\cal G}_{T}\left( E_T + {\cal Q}_T(\eta_T)\right)
$$
where ${\cal G}_{T}$ is the inverse of ${\cal L}_{T}$ acting on 
${\mathcal C}^{k+2,\alpha}_{\delta}(\ST)$.
We will find a solution to this equation in a small ball about the
origin in ${\mathcal C}^{k+2,\alpha}_{\delta}(\ST)$ for $\delta\in
(0,1)$.  
To determine the appropriate size of the ball in which to work, note 
that from Proposition \ref{pr:errest}, for any $\delta\in (0,1)$, we 
have 
\begin{equation}
\|E_T\|_{k,\alpha,\delta} \leq C\,
e^{-T/4}
\label{eq:w-errest}
\end{equation}
for a constant $C$ which is independent of $T$.
This suggests the following
\begin{definition} Fix $k\in\N$, $\alpha\in (0,1)$ and $\delta\in (0,1)$,
and let $\nu\in\R^+$. Then we define 
\[
{\mathbb B}_{\nu} = \{u \in {\mathcal C}^{k+2,\alpha}_\delta(\ST): 
\|u\|_{k+2,\alpha,\delta} \leq \nu\, e^{-T/4}\}.
\]
\end{definition}
Thus, for any $\eta\in {\mathbb B}_{\nu}$, we have
\[
|\eta|\leq\nu\, e^{-T/4} \quad \mbox{on}\quad \ST\setminus C_T
\qquad \mbox{and} \qquad |\eta|\leq \nu\,e^{-\delta T/4}\pt
\quad \mbox{on}\quad C_T.
\]
In particular, when $T$ is sufficiently large, the quadratic estimate 
(\ref{quad-est}) is valid in  ${\mathbb B}_{\nu}$.
Using Proposition \ref{pr:3} together with (\ref{eq:w-errest}) we have 
\begin{equation}
\|{\cal G}_{T} (E_T)\|_{k+2,\alpha,\delta} \leq C M\,e^{-T/4}
\label{eq:3.1.2}
\end{equation}
where $M$ is the bound on ${\cal G}_T$ acting on  
${\mathcal C}^{k,\alpha}_{\delta}(\ST)$.
In light of this we choose $\nu = 2CM$ so that we have 
${\cal G}_{T}(E_T)\subset {\mathbb B}_{\nu/2}$.
The following Lemma follows easily from the fact that we have the 
quadratic estimate above on ${\cal Q}_T(\eta)$ for 
$\eta\in {\mathbb B}_{\nu}$.
\begin{lemma}
For $\nu=2CM$ as above and $T$ sufficiently large  we have   
for any $\eta_1, \eta_2 \in {\mathbb B}_\nu$ 
\begin{equation}
\|{\cal Q}_T(\eta_1)-{\cal Q}_T(\eta_2)\|_{k+2,\alpha,\delta}\leq \frac{1}{2M} 
\|\eta_1-\eta_2\|_{k+2,\alpha,\delta}.
\label{eq:5.3}
\end{equation}
\end{lemma}
We can now establish the following
\begin{proposition}  For $\nu$ as above and $T$ sufficiently large the map 
$$
\eta \longmapsto {\cal T}(\eta)\equiv -{\cal G}_{T}\left( 
E_T + {\cal Q}_T(\eta)\right)
$$
is a contraction mapping on ${\mathbb B}_{\nu}\subset 
{\mathcal C}^{k+2,\alpha}_{\delta}(\ST)$.
\label{thm:5.1}
\end{proposition}
{\bf Proof:} Let $\eta\in{\mathbb B}_\nu$ be given.
Taking $\eta_1=\eta$ and $\eta_2 =0$ in (\ref{eq:5.3}) we see that
$-{\cal G}_{T}({\cal Q}_T(\eta))\in {\mathbb B}_{\nu/2}$.
This, together with our choice of $\nu$ shows that ${\cal T}:{\mathbb
B}_\nu\rightarrow{\mathbb B}_\nu$ as required.
Moreover for $\eta_1, \eta_2 \in {\mathbb B}_\nu$ we have
\begin{eqnarray*}
\|{\cal T}(\eta_1)-{\cal T}(\eta_2)\|_{k+2,\alpha,\delta} &=&
\|{\cal G}_{T}\left( 
E_T + {\cal Q}_T(\eta_1)\right)-{\cal G}_{T}\left( 
E_T + {\cal Q}_T(\eta_2)\right)\|_{k+2,\alpha,\delta} \\
&\leq& M\|{\cal Q}_T(\eta_1)-{\cal Q}_T(\eta_2)\|_{k+2,\alpha,\delta} \\
&\leq&\frac{1}{2} \|\eta_1-\eta_2\|_{k+2,\alpha,\delta}.
\end{eqnarray*}
This completes the proof of Proposition \ref{thm:5.1}.
\hfill$\Box$\break
Theorem \ref{thm:1} from \S \ref{se:1.2} is now a direct consequence of 
the following. 
\begin{theorem} The Lichnerowicz equation (\ref{eq:5.1}) has a unique solution
$\tpt=\pt+\eta_T$ with $\eta_T\in {\mathbb B}_{\nu}\subset 
{\mathcal C}^{k,\alpha}_{\delta}(\ST)$.
\end{theorem}
\noindent
{\bf Proof:} This follows immediately from the contraction mapping
theorem, together with Proposition \ref{thm:5.1}. 
\hfill$\Box$ \break

\section{Modifications when $\Sigma$ is complete but noncompact}
\label{se:7}

Thus far we have assumed that the three-manifold $\Sigma$ is closed.
However, it is also interesting from a physical perspectical to
consider gluing solutions of the constraint equations on various
classes of complete manifolds. Of particular interest are the two
cases when $(\Sigma,\gamma)$ is either asymptotically Euclidean
or asymptotically hyperbolic. For each of these cases, there are
well-established results concerning the existence of CMC solutions
of the Einstein constraint equation \cite{Ca1}, \cite{CIY}, \cite{AC},
\cite{ACF}. In this section we show that with only minor modifications,
the gluing constructions developed here may be extended to these
other settings. Since most of the modifications are common to both cases,
we begin with a more general discussion which describes the alterations
and new ingredients needed to extend the arguments, modulo a few
specific analytic facts. We follow this with precise statements of
definitions and facts needed for these two particular cases, along
with appropriate references.

\subsection{General preamble}

Looking back, there are only a few places in the constructions and 
arguments of the preceding sections where truly global issues arise; 
elsewhere the argument localizes in the neck region.
Specifically, the global structure of $\Sigma$ enters the analysis 
only when establishing the existence of the inverses $G_T$ and $\calG_T$ 
for the vector Laplacian $L_T$ and the linearized Lichnerowicz operator 
$\calL_T$, and also, more mildly, when determining the 
dependence on $T$ of the norms of these inverses. 

Thus suppose that $(\Sigma,\gamma)$ is a complete, noncompact, 
three-manifold, and $(\gamma,\Pi)$ is an initial data set on $\Sigma$ 
with $\mbox{tr\,}\Pi = \tau$ constant. In the noncompact setting we do 
not require that $\Pi\not\equiv 0$.  Fix points $p_1,p_2\in\Sigma$ and 
balls $B_j \ni p_j$ and construct the approximate solutions $(\gamma_T,
\mu_T,\psi_T)$ as in \S 2. Then proceed to the main part of the argument,
where first $\mu_T$ is modified to be transverse-traceless with respect 
to $\gamma_T$ using the operator $L_T$, and then $\psi_T$ is modified 
so as to be a solution of the Lichnerowicz equation. 

The function spaces we shall use in the two cases of interest are slightly 
different, and so we phrase things more abstractly in this preamble.
Thus suppose that there are Banach spaces $E$ and $F$ of vector fields and 
$E'$ and $F'$ of functions -- these will be familiar weighted H\"older 
spaces in practice -- such that the mappings 
\begin{equation}
L: E \longrightarrow F, \qquad \calL: E' \longrightarrow F'
\label{eq:miss}
\end{equation}
for the vector Laplacian $L$ and the linearized Lichnerowicz operator $\calL$ 
on $(\Sigma,\gamma)$ are both isomorphisms.  

\begin{remark} In the compact
case, we assumed that there were no conformal Killing fields vanishing 
at both $p_1$ and $p_2$, and this was enough to give the invertibility
and bounds for the inverse as $T \to \infty$ for $L_T$; likewise, the 
assumption that $\Pi \not\equiv 0$ was used to prove invertibility 
of $\calL_T$. In both the asymptotically Euclidean and asymptotically
hyperbolic settings, neither assumption is required because we shall
use spaces of functions and vector fields which decay at infinity,
and for these it is not difficult to obtain invertibility for $L_T$
and $\calL_T$ directly. Thus the restriction that the maps (\ref{eq:miss})
are both isomorphisms is not too limiting.
\end{remark}

We must also assume that the vector fields $X \in E$ decay sufficiently so
that the integration by parts 
\begin{equation}
\langle LX, X \rangle = \langle \calD X, \calD X \rangle
\label{eq:ibp}
\end{equation}
is valid for all $X \in E$. 

Since we shall always use function spaces which are localizable, i.e.\  
preserved by multiplication by functions in ${\mathcal C}^\infty_0$, 
we may define new function spaces on $\Sigma_T$ by patching together 
vector fields or functions in any of these spaces with vector fields
or functions in the appropriate H\"older spaces on $C_T$ (including functions
measured with the weight factor $w^\delta$ as required in Propositions 
7 and 8). We label these spaces $E(\Sigma_T)$, $F(\Sigma_T)$, 
$E'_\delta(\Sigma_T)$ and $F'_\delta(\Sigma_T)$. It is clear that,
presuming as above that $L$ and $\calL$ are isomorphisms, then
\[
L_T: E(\ST) \longrightarrow F(\ST), \qquad \calL_T: E'_\delta(\ST) 
\longrightarrow F'_\delta(\ST)
\]
are Fredholm for any $T$. For simplicity, we shall usually drop
the $\ST$ from this notation. 
\begin{proposition} The operator $L_T: E \longrightarrow F$
is invertible when $T$ is sufficiently large. The norm of its inverse 
$G_T$, as a mapping $F \to E$, is bounded by $C T^3$ as $T \to \infty$.
\label{pr:bdvlnc}
\end{proposition}
\begin{proposition} When $\delta \in (-1,1)$, the operator 
$\calL_T: E'_\delta \longrightarrow F'_\delta$ is invertible,
with inverse $\calG_T$ uniformly bounded as a mapping $F'_\delta
\to E'_\delta$ as $T \to \infty$. 
\label{pr:bdllnc}
\end{proposition}

We shall prove only the first of these two results; the second is proved 
in exactly the same way, but in fact is even easier to verify.

\smallskip

\noindent {\bf Proof of Proposition~\ref{pr:bdvlnc}:} Let us suppose
the result is false, so there is a sequence of values $T_j \to \infty$,
and vectors $X_j$ and $Y_j$ such that $L_{T_j} X_j = Y_j$, but with
$||X_j||_E \equiv 1$ and $||Y_j||_F = o(T^{-3})$. Also, write
$E(\Sigma_{T_j})$ and $F(\Sigma_{T_j})$ simply as $E$ and $F$. 

Following the argument of Theorem~\ref{thm:egnval}, let $q_j$ denote a 
point where $||X_j||_E$ attains its maximum. There are three cases to
consider: the first is when $q_j$ lies in $C_{T_j}$ and the norm
of $X_j$ on any compact set of $\Sigma^*$ tends to zero, the
second is when $q_j$ remains in some fixed compact set of $\Sigma^*$,
and the third is when $q_j$ leaves every compact set of $\Sigma$.
In the first two cases, the argument proceeds exactly as in \S 3.2
and \S 3.3; either we produce in the limit an `illegal' solution
$X$ on the complete cylinder, or else we produce a nontrivial vector
field $X$ on $\Sigma^*$ such that $L_c X = 0$ there. Here $L_c$ is the 
vector Laplacian for the metric $\gamma_c$ on $\Sigma^*$ which agrees
with $\gamma$ outside of the balls $B_\ell$, $\ell = 1,2$, and 
has asymptotically cylindrical 
ends near these points. This solution $X$ is bounded along these cylindrical
ends and it lies in the space $E$ on $\Sigma^* \setminus (B_1 \cup B_2)$. 
In order to carry out the usual integration by parts, to conclude that
$\calD_{\gamma_c} X = 0$, we use the analysis from \S 3.2 and the 
assumption (\ref{eq:ibp}) for $X \in E$. Hence $X$ is conformal Killing,
first with respect to $\gamma_c$, and hence with respect to $\gamma$,
on $\Sigma^*$. As in \S 3.2, it extends to a smooth solution across
the marked points $p_\ell$, and therefore gives a nontrivial element $X \in E$ 
of the nullspace of $L$ on all of $\Sigma$, which is a contradiction.
Thus it suffices to handle the third case.

Suppose then that $|X_j|$ decays to zero on every compact set of $\Sigma$.
Let $\chi$ be a smooth cutoff function which vanishes near the points
$p_\ell$ and equals one outside the balls $B_\ell$. Then $\tilde{X_j} =
\chi X_j$ satisfies $L_{T_j}\tilde{X_j} = \chi Y_j + [L_{T_j}, \chi]X_j
\equiv Z_j$. Both terms on the right decay to zero in the space $F$ on 
$\Sigma$ (not just $\Sigma^*$), whereas clearly $||\tilde{X_j}||_E \geq c 
> 0$. However, these vector fields are all defined on $(\Sigma,\gamma)$, 
and $GZ_j = X_j$. Since $G$ is a bounded operator, this 
is again a contradiction.  \hfill $\Box$

\subsection{Asymptotically Euclidean initial data}
\label{se:7.1}

We now specialize the discussion above to the case in which $(\Sigma,\gamma)$ 
is asymptotically Euclidean and $(\gamma,\Pi)$ is an appropriate data set 
on this manifold.

Recall that a metric $\gamma$ is said to be asymptotically Euclidean
(AE for short) if each end of $\Sigma$ is diffeomorphic to the exterior 
of a ball in $\R^3$, and in the induced coordinates the metric $\gamma$ 
decays to the Euclidean metric at some rate. 
More precisely, we require first that 
\[
\Phi: \Sigma \setminus K \approx \bigsqcup (\R^3 \setminus B_{R_j}),
\]
where the disjoint union on the right is finite. For simplicity we 
assume that $\Sigma$ has only one end. Then, we assume that in the 
naturally induced Euclidean coordinates $z$,
\[
|\gamma_{ij} - \delta_{ij}| \leq C|z|^{-\nu},
\]
along with appropriate decay of the derivatives.  

To formalize this, suppose that $\Sigma\setminus K \approx
\R^3 \setminus B_{R_0}$, and then define
\begin{itemize}
\item
\[
\Lambda^{0,\alpha}_{\mbox{\tiny AE}}(\Sigma) = 
\left\{u: \sup |u| < \infty, ||u||_{0,\alpha,K} < \infty, \right.  
\qquad \qquad \qquad
\]
\[
\qquad \qquad \qquad \left. \sup_{R\geq R_0}
\sup_{z \neq z' \atop R \leq |z|,|z'| \leq 2R}
\frac{|u(z)-u(z')|R^\alpha}{|z-z'|^\alpha}<\infty \right\}
\]
\item
\[
\Lambda^{k,\alpha}_{\af}(\Sigma) = 
\{u: ||u||_{k,\alpha,K} < \infty, 
\qquad (1+|z|^2)^{\frac{|\beta|}{2}}\,\del_z^{\beta} u \in \Lambda^{0,\alpha}_{\af},
\ \ \mbox{for}\ \ |\beta|\leq k\},
\]
\item
\[
r^{-\nu}\Lambda^{k,\alpha}_{\af}(\Sigma) = \{u \in \Lambda^{k,\alpha}_{\af}:
u = r^{-\nu}v \ \ \mbox{on}\ \ \Sigma\setminus K\ \ \mbox{and}\ \ 
v \in \Lambda^{k,\alpha}_{\af}\ \ \mbox{there}\}
\]
\end{itemize}
Thus, roughly speaking, a function $u$ is in $r^{-\nu}\Lambda^{k,\alpha}_{\af}$
provided it is in the ordinary H\"older space $\Lambda^{k,\alpha}$
on any fixed compact set in $\Sigma$, and if $|\del_z^\beta u| \leq
C_\beta |z|^{-\nu-|\beta|}$ along the end(s), for $|\beta| \leq k$,
with an appropriate condition for the H\"older derivatives of order
$k+\alpha$. 

We now define
\[
{\mathcal M}^{k,\alpha}_{-\nu}(\Sigma) = \{\mbox{metrics}\ \ \gamma:
\gamma_{ij}-\delta_{ij} \in r^{-\nu}\Lambda^{k,\alpha}_{\af}\}.
\]
(We are abusing notation slightly here since the metric 
coefficients $\gamma_{ij}$ are only defined on the end $\Sigma \setminus K$.)
Such metrics have been much studied 
in relativity, and \cite{Bar} collects a number of 
analytic results which are appropriate for the 
study of the Laplacians and other natural geometric operators associated 
to these metrics. 

In this AE setting it is natural to focus on initial data sets which 
would evolve into asymptotically Minkowski spacetimes, and so we 
assume that $\tau = 0$. In addition, since 
\[
\gamma \in {\mathcal M}^{k+2,\alpha}_{-\nu}(\Sigma) 
\Longrightarrow \mbox{Riem}_\gamma \in 
r^{-\nu-2}\Lambda^{k,\alpha}_{\af}(\Sigma),
\]
the Hamiltonian constraint equation (\ref{eqn2}) suggests that we
assume that
\[
\mu \in r^{-\frac{\nu}{2} - 1}\Lambda^{k,\alpha}_{\af}(\Sigma).
\]

There are many other possible choices of function
spaces in which to work. The most customary ones in the literature 
surrounding this problem are weighted Sobolev spaces. We could 
certainly have used these here too, but because weighted H\"older spaces 
have been used everywhere else in this paper it seems more natural to 
use spaces of this type here as well. 

We now define the class of elliptic AE operators. Suppose that
$P$ is an elliptic (second order) operator on $\Sigma$; we shall
assume that there is a constant coefficient elliptic operator $P_0$ of
order $2$ on $\Sigma\setminus K \approx \R^3 \setminus B_{R_0}$ such that 
\[
P = P_0 + \sum_{|\beta|\leq 2}a_\beta(z)\del_z^\beta,
\qquad \mbox{where}\quad a_\beta \in r^{-\nu}\Lambda^{k+2-|\beta|,
\alpha}_{\af}.
\]
Thus by construction, if $P$ is of this type, then
\begin{equation}
P: r^{-\nu}\Lambda^{k+2,\alpha}_{\af}(\Sigma) \longrightarrow
r^{-\nu-2}\Lambda^{k,\alpha}_{\af}(\Sigma)
\label{eq:mPae}
\end{equation}
is bounded. There are additional conditions required to ensure that this 
mapping is Fredholm. These are determined by the indicial roots:
the number $\lambda$ is an indicial root of $P$ (on one end of $\Sigma$)
if there exists a function $\phi(\theta)$, $\theta = z/|z|$, such
that $P(|z|^{\lambda}\phi(\theta)) = {\mathcal O}(|z|^{\lambda - 2-\nu})$.
Notice that no matter the value of $\lambda$ and choice of $\phi$,
the right side is always of the form $P_0(r^{-\nu}\phi(\theta)) + 
{\mathcal O}(|z|^{\lambda-2-\nu})$, where the first term here is 
homogeneous of degree $\lambda - 2$. Hence indicial roots are determined
solely by the `principal part' $P_0$ of $P$. Continuing, if we express 
$r^2P_0$ in polar coordinates as 
\[
r^2 P_0 = \sum_{j+|\beta'|\leq 2}a_{j,\beta'}(r\del_r)^j(\del_\theta)^{\beta'},
\]
(where the coefficients $a_{j,\beta'}$ are constant) then $\lambda$ is an 
indicial root for $P_0$ and hence for $P$ if and only if 
\[
\sum_{j+|\beta'|\leq 2} a_{j,\beta'} \lambda^j\del_\theta^{\beta'}\phi(\theta)
=0
\]
for some function $\phi(\theta)$. 
In particular, there is a denumerable set of indicial roots $\{\lambda_j\}$,
and these are related to eigenvalues of the `angular part' of the operator 
$P_0$. 

It is known \cite{LM}, \cite{Bar} that if $-\nu \notin 
\{\mbox{Re\,}\lambda_j\}$, then the mapping (\ref{eq:mPae}) is Fredholm 
(whereas if $-\nu$ is the real part of an indicial root, then this mapping 
does not have closed range).  We are concerned with the question of
whether this mapping is an isomorphism, or at least surjective, and if
so, for which range of weights, for the particular operators $L$ and
$\calL$. It is clear from the asymptotic flatness of the data that 
the principal parts of $L$ and $\calL$ are
\[
L_0 = -\frac12 \Delta + \frac16 \nabla \mbox{div} \qquad
\mbox{and}\qquad
\calL_0 = \Delta,
\]
respectively. (The Laplacian on the right in this expression for $L_0$
is the Laplacian on vector fields on $\R^3$, which is the same as the 
scalar Laplacian applied componentwise with respect to the standard
Cartesian decomposition.) From here we compute that
\[
\bullet \quad \mbox{the set of indicial roots for both $L$ and 
$\calL$ is}\ \ {\mathbb Z}.
\]
For the scalar Laplacian this is well known and derives ultimately
from the fact that the global temperate solutions of $\calL_0 u = 0$
are polynomials, and these have integer rates of growth. The negative
indicial roots arise from considerations of duality between weighted
$L^2$ spaces. Exactly the same
reasoning may be applied to determine the indicial roots of $L_0$,
for it is easy to see using the Fourier transform that the only global
temperate vector fields which satisfy $L_0 X = 0$ on $\R^3$ must have
polynomial coefficients. 

Then we have the following
\begin{proposition} When $P$ is equal to either $L$ or $\calL$, the 
mapping (\ref{eq:mPae}) is Fredholm of index zero when $-1 < -\nu < 0$.
\end{proposition}
We remark that the fact that these mappings are Fredholm follows from
the results quoted above and the fact that when $-\nu$ is in this
range it is not an indicial root of either operator; the fact that they
have index zero can be derived from the symmetry of these operators.
When $-\nu$ is not in this range, and is not an indicial root, then
the mapping (\ref{eq:mPae}) is still Fredholm, but no longer of
index zero, hence cannot be an isomorphism.
\begin{proposition} For $P=L$ or $\calL$ and $-1 < -\nu < 0$, the 
mappings in (\ref{eq:mPae}) are isomorphisms. 
\end{proposition}
\noindent{\bf Proof:} It suffices to show that neither $L$ nor $\calL$ 
have elements in their nullspace which decay like $r^{-\nu}$ as
$r \to \infty$. For the linearized Lichnerowicz operator $\calL$ this 
follows directly from the maximum principle. For the vector Laplacian
$L$, we first note that if
$X \in r^{-\nu}\Lambda^{2,\alpha}_{\af}(\Sigma)$ with $-1 < -\nu < 0$,
then the boundary term in the integration by parts $\langle LX,X\rangle 
= ||\calD X||^2$ vanishes, and so $\calD X = 0$. We then use a result 
of Christodoulou \cite{CCB} which says that there are 
no conformal Killing vector fields on 
AE manifolds which decay at infinity. (In fact, Christodoulou's theorem
is stated only for AE manifolds which are globally diffeomorphic
to $\R^3$, but one readily verifies from his proof that the result extends
to general AE manifolds). \hfill $\Box$

In summary, we have proved
\begin{theorem} Let $(\Sigma,\gamma,\Pi)$ be an asymptotically
Euclidean initial data set, as described in this section; in
particular, $\tau = \mbox{tr}_\gamma(\Pi) = 0$ on $\Sigma$. Choose any
two points $p_1, p_2\in \Sigma$, and suppose that if either of these
points lies on a compact (closed) component of $\Sigma$, then
there does not exist a conformal Killing field $X$ on that component which 
vanishes at that point, and furthermore, $\Pi\not\equiv 0$ on that 
component. Then there exists a one-parameter family of asymptotically
Euclidean CMC initial data sets $(\Gamma_T,\Pi_T)$ on $\Sigma_T$
satisfying the same estimates as in Theorem {\ref{thm:1}}.
\label{th:AE}
\end{theorem}  

\subsection{Asymptotically hyperboloidal initial data} 
\label{se:7.2}

Let us now turn to the case where $(\Sigma,\gamma)$ is asymptotically 
hyperbolic (or hyperboloidal, which is the terminology favored in the
physics literature). We work within the framework of conformally 
compact manifolds, which is quite natural from Penrose's formulation.
This is explained carefully in the introduction to \cite{AC}. 

Recall that a metric $\gamma$ on the interior of a manifold with boundary 
$\overline{\Sigma}$ is said to be conformally compact if $\gamma = 
\rho^{-2}\olg$, 
where $\olg$ is a nondegenerate metric on the closure $\overline{\Sigma}$ and 
$\rho$ is a defining function for $\del \Sigma$, i.e.\ $\rho$ is nonnegative 
and vanishes precisely on the boundary, with nonvanishing differential. We 
shall assume that $\olg$ and $\rho$ are both smooth, or more
accurately, polyhomogeneous (i.e. having complete expansions involving 
powers of a smooth defining function as well as powers of its logarithm); 
this latter 
assumption is most reasonable because it occurs generically \cite{AC}. 
Though we shall not do so, it is also possible to assume that $\olg$ and 
$\rho$ are in ${\mathcal C}^{k+2,\alpha}(\overline{\Sigma})$, and only minor and 
obvious perturbation arguments  are needed to handle this more general case.

The metric $\gamma$ is complete on the interior $\Sigma$ of 
$\overline{\Sigma}$; moreover, assuming that $|d\rho|_{\olg} = 1$ at $\rho=0$, 
then the curvature tensor $\mbox{Riem}_g$ is asymptotically isotropic
as $\rho \to 0$, with sectional curvatures tending to $-1$. We call any
such space AH (standing for either asymptotically hyperbolic or 
asymptotically hyperboloidal, depending on the reader's preference).
As for the tensor $\Pi = \mu + \frac13 \tau \gamma$, we assume that 
$\tau$ is constant, of course, and also that the tensor coefficients
$\mu_{ij} = {\mathcal O}(\rho^{-1})$, or equivalently $\mu^{ij} = 
{\mathcal O}(\rho^3)$. (In either
case, we are writing the components of the tensor $\mu$ relative to 
a smooth background coordinate system on $\overline{\Sigma}$.)
Assuming that $\tau = 3$ and with the curvature normalization above,
the constraint equations become
\[
\mbox{div\,}_\gamma\,(\mu) = 0, \qquad R + 6 = |\mu|^2.
\]
(Recall that $|\mu|^2 = {\mathcal O}(\rho^2)$.) 

The existence of AH initial data sets $(\gamma,\Pi)$ satisfying these
hypotheses, as well as the polyhomogeneous regularity of this data
(assuming that $\olg$ is smooth) is the main topic of \cite{AC}. 
The special case of these results, in which one assumes that $\Pi$ is 
pure trace, is already handled in \cite{ACF}. Some of the main tools 
required here are based on the linear analysis of geometric elliptic 
operators on conformally compact spaces as developed in \cite{Ma1}, 
although the authors of \cite{AC} and \cite{ACF} base their work on the 
somewhat later and less precise, although simpler 
methods of \cite{A}, cf. also \cite{L} for a sharper version of these 
methods. The regularity theory of \cite{AC} is very closely related to the
results of \cite{Ma2}.

Quoting the main result of \cite{AC}, we assume that $\Sigma$ is 
endowed with an AH initial data set $(\gamma,\Pi)$, where $\gamma$ and 
$\mu$ are polyhomogeneous. That is, with $\olg$ a smooth metric on 
$\overline{\Sigma}$ and $x$ a smooth defining function for $\del 
\Sigma$, we have
\[
\rho \sim x + \sum_{k\geq 0}\sum_{\ell=0}^{N_k} a_{k\ell}(\log x)^k x^\ell,
\]
and
\[
\mu_{ij} \sim \mu^{(-1)}_{ij}x^{-1} + \sum_{k\geq 0}\sum_{\ell=0}^{N_k} 
\mu_{ij}^{(k\ell)} (\log x)^k 
x^\ell,
\]
where all coefficients $a_{k\ell}$, $\mu^{(-1)}_{ij}$, 
$\mu_{ij}^{(k\ell)}$ are smooth on 
$\overline{\Sigma}$
and for each $j$, vanish when $k > N_j$. (In addition, $a_{2\ell} = 0$
for $\ell>0$.)

We require the mapping properties of the vector Laplacian 
$L = -\mbox{div} \circ \calD$ and the linearized Lichnerowicz operator 
$\calL = \Delta - |\mu|^2 - 3$ associated to this data.  These follow 
directly from the general theory of \cite{Ma1}, cf.\ also \cite{L}. 

Choosing local coordinates $(x,y)$ near $\del\Sigma$, then recall from these 
papers that any operator of the form
\[
P = \sum_{j+|\alpha|\leq 2} a_{j\beta}(x,y)(x\del_x)^j (x\del_y)^\alpha
\]
with coefficients $a_{j\beta}$ smooth (or just polyhomogeneous)
is said to be uniformly degenerate (of order $2$). Its `uniformly 
degenerate symbol' is defined to be 
\[
\sigma_2(P)(x,y;\xi,\eta) = \sum_{j+|\beta| =2} a_{j\beta}(x,y)\xi^j
\eta^\beta,
\]
and $P$ is elliptic in this calculus if this symbol is nonvanishing
(or an invertible matrix, if $P$ is a system) when $(\xi,\eta) \neq 0$. 
It is not hard to check that both of the operators $L$ and $\calL$
above are elliptic uniformly degenerate operators. 

The function spaces on which we let these operators act are the weighted
H\"older spaces $x^\nu \Lambda^{k,\alpha}_{\ah}(\Sigma)$. These are taken to be
the usual H\"older spaces on a compact subset of $\Sigma$ and, in a 
boundary coordinate chart are defined as follows:
\begin{itemize}
\item
\[
\Lambda^{0,\alpha}_{\ah}(\Sigma) = \left\{u: \sup |u| < \infty, \quad
\sup \frac{|u(x,y) -u(x',y')|(x+x')^\alpha}{(|x-x'|+|y-y'|)^\alpha} < 
\infty\right\},
\]
\item
\[
\Lambda^{k,\alpha}_{\ah}(\Sigma) = \{u: 
(x\del_x)^j(x\del_y)^\beta u
\in \Lambda^{0,\alpha}_{\ah},\quad j+|\beta|\leq k\},
\]
\item
\[
x^\nu\Lambda^{k,\alpha}_{\ah}(\Sigma) = \{u = x^\nu v: v \in 
\Lambda^{k,\alpha}_{\ah}\}.
\]
\end{itemize}
Clearly
\[
P: x^\nu \Lambda^{k+2,\alpha}_{\ah}(\Sigma) \longrightarrow
x^\nu \Lambda^{k,\alpha}_{\ah}(\Sigma)
\]
is bounded for any $\nu$, but just as in the AE case, restrictions on
$\nu$ must be imposed to ensure that this map is Fredholm.

To state these, we say that a number $\lambda$ is an indicial 
root for $P$ if $P x^\lambda = {\mathcal O}(x^{\lambda + 1})$ as $x \to 0$. 
It is always true that $P x^\lambda = {\mathcal O}(x^\lambda)$ and the
indicial root condition requires that $\lambda$ satisfy a polynomial 
equation involving 
the coefficients $a_{j0}(0,y)$. (It is possible that these roots could 
depend on $y$, but in our application this does not occur.) A brief
calculation, using that $|\mu|^2 \to 0$ as $x \to 0$, gives:
\begin{itemize}
\item The indicial roots for $L$ are $0$ and $4$,
\item The indicial roots for $\calL$ are $-1$ and $3$.
\end{itemize}
We now state one of the main theorems from \cite{Ma1}, specialized
to this particular setting:
\begin{proposition} Suppose that $0 < \nu < 4$ and $-1 < \nu' < 3$.
Then the mappings
\[
L: x^\nu \Lambda^{k+2,\alpha}_{\ah}(\Sigma) \longrightarrow
x^\nu \Lambda^{k,\alpha}_{\ah}(\Sigma)
\]
and
\[
\calL: x^{\nu'} \Lambda^{k+2,\alpha}_{\ah}(\Sigma) \longrightarrow
x^{\nu'} \Lambda^{k,\alpha}_{\ah}(\Sigma)
\]
are Fredholm of index $0$. Furthermore, the nullspace of either
of these operators is independent of $\nu \in (0,4)$ and $\nu'\in
(-1,3)$.
\label{pr:ahfr}
\end{proposition}

To conclude that these maps are isomorphisms, it suffices to prove
\begin{proposition}
For $\nu$ and $\nu'$ in the ranges stated in the previous proposition,
the maps $L$ and $\calL$ are injective.
\end{proposition}
\noindent{\bf Proof:} For $\calL$ this is quite easy. 
In fact, if $\calL \phi = 0$, then $|\phi| \leq C x^{\nu'}$ for 
any $\nu' \in (-1,3)$, and therefore $\phi \to 0$ at $\del\Sigma$.
But the terms of order $0$ in $\calL$ are strictly negative, and so
the maximum principle implies that $\phi \equiv 0$. As for $L$,
we note that if $LX = 0$, then the coefficients of $X$ are
${\mathcal O}(x^4)$, and so we may perform the usual integration 
by parts to obtain that $\calD X = 0$. By the conformal invariance
of the nullspace of $\calD = \calD_\gamma$, we also have that
$\calD_{\olg} X = 0$. Thus $X$ is a (smooth or polyhomogeneous)
conformal Killing vector field for the metric $\olg$ on $\overline{\Sigma}$, 
with coefficients vanishing to order $4$ at the boundary. Since we
already know that $X$ is polyhomogeneous, the equation $\calD_{\olg} X = 0$
now implies that $X$ vanishes to infinite order as $x \to 0$, and it
is known that no such conformal Killing field can exist \cite{COM}. 
Thus we have shown that
\begin{proposition} The mappings in Proposition~\ref{pr:ahfr} are
both isomorphisms.
\label{pr:ahis}
\end{proposition}

Therefore, we may use the spaces occuring here as the spaces
$E$, $F$, $E'$ and $F'$ needed in Propositions \ref{pr:bdvlnc}
and \ref{pr:bdllnc}. This proves
\begin{theorem} Suppose that $(\Sigma,\gamma,\Pi)$ is an asymptotically
hyperbolic initial data set, so that $\gamma$ is a conformally
compact metric and $\tau = \mbox{tr}_\gamma(\Pi) \equiv 3$.
Let $p_1, p_2 \in \Sigma$, and suppose that if either of these
points lies on a compact (closed) component of $\Sigma$, then
there does not exist a conformal Killing field $X$ on that
component which vanishes at that point. Then there exists
a one-parameter family of asymptotically hyperbolic CMC initial 
data sets $(\Gamma_T,\Pi_T)$ on the manifold $\Sigma_T$
satisfying the same estimates as in Theorem {\ref{thm:1}}.
\label{th:AH}
\end{theorem}

\section{The geometry of the neck} 
\label{se:8}

We have now shown in the three main settings ($\Sigma$ compact 
or AE or AH) that it is possible to produce one-parameter families 
of CMC initial data sets $(\Gamma_T,\Pi_T)$ which are 
arbitrarily good approximations to the original initial data
set $(\gamma,\Pi)$ away from the points $p_1$, $p_2$. As 
$T \to \infty$, various geometric quantities associated to
these tensors degenerate in the neck region. We describe this in 
more detail now.

We shall only consider quantities which depend on at most two derivatives 
of the metric, and so we fix any $k \geq 2$ and $\alpha \in (0,1)$
and construct the initial data sets as in \S\S 2--7. 

Recall the notation $\Sigma_R^* = \Sigma \setminus (B_R(p_1) \cup B_R(p_2))$.
Next, for $b \in (0,R)$, let $A_j(b) = B_R(p_j) \setminus B_b(p_j)$,
and finally denote by $C_{T,b}$ the part of the neck region not covered 
by these annuli. Then for each $T > 0$, 
\[
\Sigma_T = \Sigma_R^* \sqcup (A_1(b) \cup A_2(b)) \sqcup C_{T,b}.
\]
In either of the annuli and in the truncated neck region we 
shall use the $(s,\theta)$ coordinates, so that, for example,
$A_1(b) = \{T/2 - \beta \leq s \leq T/2\} \times S^2$ and
$C_{T,b} = \{-T/2 + \beta \leq s \leq T/2 - \beta\} \times S^2$,
for $\beta = -\log b$. Write $\psi_T^4 \gamma_T = \tgT$;
of course, this metric is the same as $\gamma$ in $\Sigma_R^*$,
and the deviation between these two metrics in either of the
half-necks $0 < |s| < T/2$ is ${\mathcal O}((e^{-T}\cosh s)^2)$. 
We shall estimate $|\Gamma_T - \tgT|_{\tgT}$ (i.e.\ the difference between
the actual solution $\Gamma_T$ and the approximate solution $\tgT$)
in each of these regions.

By definition, $\Gamma_T = (\psi_T + \eta_T)^4\gamma_T$. Hence
\[
|\Gamma_T - \tgT|_{\tgT} \approx \psi_T^{-4}((\psi_T + \eta_T)^4
-\psi_T^4) \leq C |\eta_T/\psi_T|.
\]
Recall also from \S \ref{se:6} that $|\eta_T| \leq \nu e^{-T/4}w_T^\delta$,
Furthermore, $1/2 \leq w_T/\psi_T \leq 2$ everywhere. 
Therefore:
\begin{itemize}
\item In $\Sigma_R^*$, $\tgT = \gamma$ and $\psi_T \approx 1$.
Thus 
\begin{equation}
|\Gamma_T - {\tgT}|_{\tgT} \leq C e^{-T/4}
\label{eq:est1}
\end{equation}
in this region.
\item In all of $C_{T}$, we have 
\begin{equation}
|\Gamma_T - \tgT|_{\tgT} \leq C e^{-T/4}(e^{-T/4}\cosh(s/2))^{\delta - 1}
= C e^{-\delta T/4}(\cosh(s/2))^{\delta - 1}.
\label{eq:est3}
\end{equation}
\item In particular, in $A_1(b)$, $|\eta_T| \leq \nu e^{-T/4}w_T^\delta$, 
and so
\begin{equation}
|\Gamma_T - \tgT|_{\tgT} \leq C e^{-T/4}(e^{-T/4}e^{s/2})^{\delta - 1}
= C e^{-\delta T/4}e^{(\delta-1)s/2} \leq C e^{(1-\delta)\beta/2 - T/4}.
\label{eq:est2}
\end{equation}
The estimate in $A_2(b)$ is identical.
\end{itemize}
Notice in each of these cases that since $\delta < 1$, this difference
of tensors is much smaller than $\tgT$, and therefore geometric
quantities relative to $\Gamma_T$ are very well approximated by
the same quantities relative to $\tgT$. 

The following two results are immediate corollaries.

\begin{proposition}
Let $c$ be any ${\mathcal C}^1$ path in $\Sigma_T$ which runs through 
the neck region. For simplicity we shall assume that it has the
simple form $s \mapsto (s,\theta_0)$, $-T/2 \leq s \leq 
T/2$. Then 
\[
\mbox{\rm length}_{\Gamma_T}(c) \sim \mbox{\rm length}_{\tgT}(c) + 
{\mathcal O}(e^{-\delta T/4}).
\]
In particular, for the portion of $c$ which lies in either of the annuli
$A_j(b)$, we may substitute $\mbox{\rm length}_{\gamma}(c)$ for
$\mbox{\rm length}_{\tgT}(c)$ on the right side here.
\end{proposition}
\begin{proposition}
In $C_{T,b}$, the full curvature tensor $\mbox{\rm Riem}_{\Gamma_T}$ satisfies
\[
|\mbox{\rm Riem}_{\,\Gamma_T}|_{\tgT} \leq C \psi_T^{-4} \leq  
C e^{T}(\cosh s)^{-2}.
\]
In particular, the scalar curvature $R_{\Gamma_T}$ satisfies this
same estimate.
\end{proposition}

\noindent Of course, we could also compute finer asymptotics for the
full curvature tensor and for the scalar curvature function.

Estimates for the extrinsic curvature tensor are also easy to
obtain. We have already shown that the deviation of the trace-free
part of $\Pi_T$ from that of $\Pi$ on $\Sigma_R^*$ is 
bounded by $T^3e^{-3T/2}$. On the other hand, relying on 
the fact that $|\mu|_{\gamma_c} \leq Ce^{-T/2}\cosh s$ together with the
preceding estimates, we see that
\[
|\Pi_T|_{\Gamma_T} \leq C e^{T/2}(\cosh s)^{-1}, \qquad
\mbox{on}\qquad C_T.
\]
We may use this as a check on this last proposition: the constraint
equation (\ref{eqn2}) gives the scalar curvature in terms of
$|\Pi_T|_{\Gamma_T}^2$, and we see that the estimates are the 
same. However, we also see from this that if the original tensor
$\mu = 0$ in both balls $B_1$ and $B_2$, then $|\Pi_T|_{\Gamma_T}$ is
bounded in $C_T$, and hence $R_{\Gamma_T}$ is also. 

\section{Examples and applications}
\label{se:9}

There are a number of constructions and applications of special interest
in general relativity to which our results pertain. In this 
section we briefly describe several of these. This discussion is 
intended to be informal and descriptive, and so we do not present 
these results formally, but given the rest of the contents of this 
paper, it should be obvious how to do so. These examples are 
separated into different subsections, but these divisions
are artificial and are intended for guidance only.

\medskip

\vfill \eject

\noindent{\bf Adding wormholes}

\medskip

The first construction is that of adding a physical wormhole to 
(almost) any globally hyperbolic solution of the vacuum 
(Lorentzian) Einstein equations which admits a CMC Cauchy surface. 
One proceeds by: a) choosing such a Cauchy surface $\Sigma$;
b) choosing any pair of points on $\Sigma$; c) verifying the
conditions of Theorems \ref{thm:1}, \ref{th:AE} or \ref{th:AH};
d) carrying out the gluing construction as outlined in \S\S 2--6; and 
finally, e) evolving the resulting initial data. Standard well-posedness 
theorems \cite{CB}, \cite{CBGB} guarantee that there is a unique (up to 
spacetime diffeomorphism) maximal
spacetime development of this initial data set. In other words, there is a
spacetime which satisfies the vacuum Einstein equations, has the topology
$\Sigma_T \times \R$, restricts to the wormhole initial data on $\Sigma_T 
\times \{0\}$, and also contains every other spacetime which satisfies these
three properties. Of course, this spacetime will probably not last very 
long, and indeed it is likely that the evolving wormhole will pinch off 
quickly, thus preventing the spacetime from evolving further in time. 
Nonetheless, there is a spacetime 
solution which evolves from this initial data and which therefore
contains a spacetime wormhole connecting the chosen points;
in addition, the evolving spacetime remains essentially unchanged
from its previous unglued state outside the domain of influence of 
the initial data on the wormhole region. 

There do exist spacetimes for which this wormhole addition cannot be 
carried out: these are the ones which do not carry any set of CMC data 
which matches the hypotheses of our theorems. A simple example of this
phenomenon is the `spatially compact Minkowski space' $T^3 \times \R$, 
since all CMC slices of this spacetime are compact and have vanishing 
extrinsic curvature, and hence the hypotheses of Theorem 1 are not satisfied.

\medskip

\noindent {\bf Multiple black hole spacetimes}

\medskip

Over 40 years ago Misner \cite{CWM} constructed a number of explicit 
initial data sets involving two asymptotically Euclidean regions 
connected by a wormhole. Soon thereafter, he and others, cf. \cite{HE}, 
recognized that since the wormhole inevitably contains a minimal 
two-sphere, and since the presence of such a two-sphere necessitates
the existence of an apparent horizon, the spacetime developments of 
these initial data sets must contain black holes. In subsequent work,
Misner \cite{CWM2} 
constructed explicit (series form) sets of initial data with more than
one wormhole; the spacetime developments of these data must then 
contain more than one black hole. Misner's data has been used extensively
by numerical relativists \cite{GC} in studying multi-black hole
spacetimes.

Our gluing constructions yield a much wider class of multi-black hole 
solutions. Most importantly, we can successively add black holes to any 
given asymptotically flat spacetime.  (We say that a spacetime is 
asymptotically flat if it is foliated by asymptotically Euclidean 
initial data sets; see chapter 11 in \cite{Wa} for a more rigorous 
definition.) The idea is as follows: start with any asymptotically 
flat spacetime which admits a maximal (i.e. zero mean curvature), 
asymptotically Euclidean hypersurface. 
Choose such a maximal Cauchy surface with AE data 
$(\Sigma,\gamma,\Pi)$ and a point on this hypersurface, and then glue 
on a copy of Euclidean space $\R^3$ with zero extrinsic curvature (which 
is a maximal data set for Minkowski space). The result is a new initial
data set with two AE regions connected by a wormhole, with the 
data largely unchanged on either side of the wormhole. As before, the 
wormhole contains a minimal two-sphere, and so black
hole formation is inevitable in the spacetime development of this data. 
We may now continue this process, taking this new data and gluing on 
a further copy of maximal data for Minkowski space. This yields a 
spacetime with two black holes. Proceeding further, we obtain spacetimes 
with an arbitrary (finite, or even infinite) number of black holes,
with each additional black hole leaving the data for the previous
black holes largely unchanged.

If one glues two AE sets of data together, the resulting data set has two
AE ends. From a mathematical point of view, there is nothing to
distinguish one of these ends from the other. However, in using Einstein
solutions to model spacetime physics, one often makes a choice of one of
the ends, and its spacetime development, to be the physically accessible
``exterior'', with the other end corresponding to a physically inaccessible
``interior''. The exterior is the home of the ``observers at infinity'' 
while the interior is somewhere ``inside the black hole''. Apart from 
considerations of physical modelling, the choice is arbitrary.

Multi-black hole spacetimes have several ends, and any one of these
may be regarded as {\it the} exterior. However, to model spacetimes 
with observable interacting black holes, one must construct multi-black hole
initial data sets with the chosen exterior asymptotic region directly 
connected to each of the wormholes. Each of these wormholes leads through
to its opposite interior end. These interiors may be independent of
one another, or they may coincide as a single asymptotic region. In 
other words, one may 
envision the manifold as having, for example, two asymptotic regions 
connected by several necks. This 
case corresponds to Misner's ``matched throat'' data
sets \cite{CWM2}, and is no more difficult to construct from
our point of view than a solution with independent interior ends.

We emphasize that the multi-black hole data sets we are describing here 
are not presented as explicitly as Misner's examples, because to obtain 
them one must solve the constraint equations in conformal form. In 
practice, there are efficient ways to do this numerically. 
As already noted, one virtue of our results is that we obtain 
solutions which leave the spacetime essentially unchanged away from 
the domain of influence of the added black hole. This feature is quite 
interesting and should be useful to numerical relativists.

\medskip

\noindent {\bf Adding AE ends to closed initial data sets}

\medskip

Every closed three manifold $\Sigma$ admits a CMC solution of the 
vacuum constraint equations: we choose $\gamma$ to be a metric of constant 
negative scalar curvature $R(\gamma)=-6 $ (these always exist), and set 
$\Pi=\gamma$; $(\Sigma, \gamma, \gamma)$ is then a solution of the 
constraint equations, with 
constant mean curvature $\tau=3$. It is of interest to know whether 
every manifold of the form $\Sigma\setminus\{p\}$ with $\Sigma$ closed
admits AE initial data sets. For a manifold $\Sigma$ which admits a metric 
of positive scalar curvature 
with no conformal Killing fields and with a nonvanishing symmetric
transverse-traceless $2$-tensor, our methods show that this is the 
case. To see this, we first use \cite{I1} to get that any such $\Sigma$ 
admits a solution $(\Sigma, \gamma, \Pi)$ of the constraint equations 
with $\tau=0$. Then by Theorem~\ref{th:AE}, we may glue the standard
AE initial data set on $\R^3$ with vanishing extrinsic curvature 
to $(\Sigma,\gamma,\Pi)$ to get an AE initial data 
set on $\Sigma\setminus\{p\}$. Note the closely related result, that when 
$\Sigma$ is compact, $\Sigma \setminus \{p\}$ admits a complete scalar 
flat metric, i.e. a time-symmetric AE initial data set, if and only if 
$\Sigma$ admits a metric of positive scalar curvature, cf.\ \cite{Ma2}. 

On the other hand, suppose that $\Sigma$ has the property that any metric 
of nonnegative scalar curvature on it is flat (for example, $\Sigma = T^3$
has this property). Then by (\ref{eqn2}), any initial data set 
$(\Sigma,\gamma,\Pi)$ with $\tau = 0$ has nonnegative scalar curvature.
By the hypothesis on $\Sigma$, this means that $\gamma$ is flat, and so,
using (\ref{eqn2}) again, we must have $\Pi \equiv 0$. But $(\Sigma,
\gamma,0)$ does not satisfy the hypotheses of Theorem~\ref{th:AE},
and so we cannot use our gluing theorem to produce AE initial data
sets on $\Sigma\setminus\{p\}$. 

We now comment briefly on the physical interpretation of the solutions we
obtain by gluing AE ends to closed initial data sets as above. Suppose
we view the compact side as the exterior, accessible region. This 
data has a minimal two-sphere as before, and therefore 
most likely evolves into a spacetime with an apparent horizon.
But the problem is that there is no clear and unambiguous definition 
of a black hole relative to an exterior spacetime region which is 
not asymptotically flat. This makes the interpretation of the 
spacetime development in this case less transparent, but it also 
makes such spacetimes rather interesting. Indeed, one
might be able to learn a lot about how to work with black holes in
spatially closed spacetimes by careful study of these examples.

As noted above, if a spacetime contains a wormhole connecting a pair of
regions, one may choose either of the pair of regions as the exterior
accessible region. The choice is mathematically arbitrary, but physically
important for modelling. Thus the  spacetimes one obtains by gluing
together data from an asymptotically flat spacetime and data from a
spatially compact spacetime, and then evolving, may be viewed either as a
spatially compact cosmology with a black hole-like object in it (with
the closed region as the exterior), or as a black hole with a 
nontrivial interior solution. The construction is the same; all we 
are doing is redefining which region we consider to be the exterior, 
physical region. The wormhole may cut off very quickly and render the 
interior solution physically irrelevant; this is the nature of black 
hole physics, and we believe that it can be studied meaningfully with 
these gluing constructions.

\medskip
\noindent {\bf Adding AH ends to closed initial data sets}
\medskip

In distinction to the AE case, since AH initial data sets have nonvanishing 
mean curvature $\tau = 3$, it is possible to glue an AH initial data set 
to an appropriately chosen intial data set on any compact manifold $\Sigma$.
In fact, as above, let $\gamma$ be a metric of constant
negative scalar curvature $R = -6$ on $\Sigma$, so that $(\Sigma,\gamma,
\gamma)$ is an initial data set also with $\tau = 3$. By Theorem~\ref{th:AH},
we may glue to it any AH initial data set (in particular, the hyperboloid 
$({\mathbb H}^3, \gamma_0, \gamma_0)$, where $\gamma_0$ is the standard 
hyperbolic metric) at any point $p \in \Sigma$. This proves in particular
that for every closed manifold $\Sigma$, the punctured manifold 
$\Sigma\setminus\{p\}$ admits an AH initial data set. In fact, we may 
iterate this procedure and obtain AH initial data sets on $\Sigma 
\setminus \{p_1, \ldots, p_k\}$ for any collection of distinct points on 
any closed $3$-manifold $\Sigma$. We note again that the analogue of 
this question for AE vacuum initial data sets is still open.  

As a simple and interesting example of this last construction, 
let $(\Sigma,\gamma)$ be a closed hyperbolic $3$-manifold. 
By adjoining to it the hyperboloid as above we obtain an AH initial 
data set on $\Sigma\setminus\{p\}$. In fact, we may find such initial
data sets which are arbitrarily small perturbations of the initial 
hyperbolic metric away from the neck. 

It is also straightforward to glue together two or more AH data sets. 
Not surprisingly, it is impossible to glue CMC AH data 
sets to CMC AE data sets because the constant mean curvatures in the 
two cases are different. 

We are left with the question of whether the spacetimes which result 
from evolving these glued AH initial data sets develop black holes. 
Again, if we view the spatially compact part of the data as the exterior, 
physically accessible, part of the solution, then this question does not 
have a definite meaning, since as before, black holes are only well-defined 
in spacetimes with asymptotically flat ends. 

\vfill\eject

\end{document}